\documentclass[structabstract]{aa}  
\usepackage{graphicx}
\usepackage{savesym}
\usepackage{amsmath}
\savesymbol{iint}
\usepackage{txfonts}
\restoresymbol{TXF}{iint}
\usepackage{natbib}
\def\grad {{\nabla}}
\begin{document}
   \title{Three-dimensional modeling of radiative disks in binaries}

   \author{G. Picogna
          \inst{1}
          \and
          F. Marzari\inst{1}
          }

   \institute{Department of Physics \& Astronomy, University of Padova,
              via Marzolo 8, I-35131 Padova\\
              \email{picogna@pd.infn.it}
             }

   \date{Received September 15, 1996; accepted March 16, 1997}

\abstract
{Circumstellar disks in binaries are perturbed 
by the companion gravity causing significant
alterations of the disk morphology. Spiral 
waves due to the companion tidal force 
also develop in
the vertical direction and affect 
the disk temperature profile. 
These effects may significantly influence 
the process of planet formation. 
}
{We perform 3D numerical simulations of disks 
in binaries with different initial dynamical configurations and 
physical parameters. Our goal is to investigate 
their evolution and their  propensity to 
grow planets.  
}
{ We use an improved version of the SPH code VINE 
modified to better account for momentum and energy conservation
via variable smoothing and softening length. The 
energy equation includes a flux--limited radiative 
transfer algorithm. The disk cooling is obtained with 
the use of ``boundary particles'' populating 
the outer surfaces of the disk and radiating to 
infinity. 
We model a system made of star/disk + star/disk where the 
secondary star (and relative disk) is less massive than 
the primary.
}
{The numerical simulations performed for different values of 
binary separation and disk density show that 
trailing spiral shock waves develop when the stars 
approach their pericenter. Strong hydraulic jumps 
occur at the shock front, in particular for 
small separation binaries, creating breaking waves, and
a consistent mass stream between the 
two disks.
Both shock waves and mass transfer
cause significant heating of the disk. At apocenter
these perturbations are reduced and the disks are cooled
down and less eccentric.
}
{The disk morphology is substantially affected by the 
companion perturbations, in particular in the vertical 
direction. The hydraulic jumps may slow
down or even halt the dust coagulation process.
The disk is significantly heated up by spiral waves and mass 
transfer and the high gas temperature may prevent the 
ice condensation by moving outward the ``snow 
line''. The disordered motion triggered by the 
spiral waves may, on the other hand, favor the direct
formation of large planetesimals from pebbles. 
The strength of the hydraulic jumps, disk heating, 
and mass exchange depends on the 
binary separation, and for larger semi-major axes,
the tidal spiral pattern is substantially 
reduced. The
environment then appears less hostile to planet formation.
}

   \maketitle
%

\section{Introduction}

   According to \cite{dunma91}, more than 50\% of G type stars
   are in multiple systems. The multiplicity appears
   to depend on the stellar mass with a higher frequency for 
   massive stars, roughly 70\% \citep{maso98}, and a lower
   frequency of about 30\% for less massive M stars (see 
   \cite{lada06} for a review of the data). It is also 
   suspected that most stars could form in binary or 
   higher multiplicity systems disintegrating at later times
   via dynamical interactions or decay \citep{rei00}.
   As for single stars, binary stars in the early stages of 
   their evolution are surrounded by protoplanetary disks.
   An example is the L1551 IRS 5 system, which contains two protostars, 
   each surrounded by a circumstellar disk \citep{rod98,oso03}.
   Spatially resolved observations of disks in binaries in the Orion nebula 
   cluster \citep{dae12} suggest that the fraction of circumstellar disks 
   around individual components of binary systems is about 40 \%, only 
   slightly lower than that for single stars (roughly 50\%). 
   This is possibly due to the impact of the companion star on the 
   disk evolution which causes disk truncation, eccentric shape, 
   warping, and heating \citep{Nelson00,Kley08,paard08,
   Marzari09,mabash2}.

   The physical properties of circumstellar disks are relevant for
   forming planetary systems. Perturbed disks, like those in close 
   binaries, may affect planet accretion at different stages, producing a 
   population of planets that differ from those around single stars. 
   The process by which dust particles evolve into kilometer--sized 
   planetesimals, which is still not fully understood for single stars 
   \citep{wei77,wurblu08}, may be altered in disks around binaries. 
   Spiral waves excited by the companion perturbations may, via Epstein
   drag coupling, affect the relative velocity of the colliding particles, 
   the drift rate towards the star and the vertical settling speed. All 
   these parameters strongly influence the collisional sticking process 
   and dust agglomeration into larger bodies, and the binary perturbations 
   appear to act against a fast dust coagulation. On the other hand,
   the large scale motions in these disks, excited by the gravitational 
   pull of the companion star, may locally favor concentration and 
   subsequent accumulation of dust and pebbles directly into planetesimals 
   \citep{johan07,cuzzi08}. Even the planetesimal accretion phase appears 
   more complex in close binary systems due to the increase in the mutual
   impact velocity between the planetesimals caused by the combined action
   of secular perturbations by the companion star and gas drag effects
   \citep{mascho,thebs04,theb06,thems08,Kley08,thems09,Marzari09,xie09,
   xie10,paard08,mabash2}. 

   In spite of all these additional complications, which seem to lower 
   the efficiency of planet formation, about 50 planets are known to date 
   in binary stars, among which are also the newly discovered planet in the 
   Alpha Centauri system \citep{dum12}. 
   In addition, according to \cite{bona} and \cite{mugra}, the frequency 
   of planets in binaries do not appear to differ much from that of planets 
   orbiting single stars.
   However, it is reasonable to expect that the differences in the two 
   initial steps of planet formation -- dust coagulation and planetesimal 
   accretion -- would have a significant impact on the final physical 
   properties and dynamical architecture of planetary systems around single 
   and binary stars.
   For example, \cite{zumaz02}, \cite{egg04}, \cite{mug05}, and \cite{desi07} 
   find some discrepancies in the mass--period and eccentricity--period 
   diagrams of planets in binaries and single stars.

   The dynamical structure of circumstellar disks in binaries is 
   crucial not only to understand the processes of planet formation but 
   also to predict the migration of planets. 
   Their morphology, temperature, and density profiles are 
   significantly affected by the excitation of eccentric modes 
   \citep{paard08,Kley08,Marzari09}, shock waves formation, and 
   mass exchange \citep{Nelson00}. These differences, compared to disks
   around single stars, may influence the migration speed and direction.
   Population synthesis calculations, like those described in 
   \cite{morda12}, have shown that differences in migration may lead to 
   distinct predictions concerning the final orbital and mass 
   distributions of planets. To understand the architecture of 
   planetary systems in binaries, in particular those with small 
   separations, it is then important to know the morphology and physical state 
   of circumstellar disks in the crucial phases of planet growth, i.e.,
   during dust coagulation and planetesimal accumulation.  

   In this paper we model the 3D evolution of circumstellar 
   disks in binaries using the SPH code VINE \citep{Wetzstein09, Nelson09}.
   The original code had been upgraded to improve both
   momentum and energy conservation.
   In addition, to better model optically thick disks 
   in their initial stages of evolution, we have implemented a radiative 
   energy equation. This is particularly recommended since the companion star
   induces strong spiral waves in the disks that may generate local strong 
   shocks and compressional heating violating the local isothermal 
   approximation. In addition, these shocks may also produce notable effects 
   in the vertical direction, potentially affecting the dust 
   coagulation process. 

   In Section~\ref{model} we focus on the upgrades to the SPH code VINE 
   and on the implementation of the radiation hydrodynamics. In 
   Section~\ref{setup} we describe the initial conditions of the 
   simulations concerning different density disks in close and wide binary 
   configurations. In Section~\ref{results} we analyze the outcomes of the 
   simulations and discuss some theoretical implications. Finally in
   Section~\ref{conclusions} we summarize our results, discuss their 
   relevance for planet formation, and comment on possible future 
   improvements in the study of disks in binaries. 
\section{The model}\label{model}

   To compute the evolution of the disks surrounding the stars in the 
   binary system, we use the hydrodynamical code VINE \citep{Wetzstein09, 
   Nelson09}. It is based on the SPH (smoothed particle hydrodynamics) 
   algorithm that solves the equations of fluid dynamics by replacing the 
   fluid with a set of particles \citep{Gingold77}.\\
   We have updated the code with some important features that improve momentum 
   and energy conservation during the simulations. In addition, we 
   have implemented a fully radiative approach so that the viscous heating is 
   diffused in the disk and emitted at the disk's outer borders. The 
   flux--limited approximation, as described in \cite{Levermore81}, is 
   used in the code. Here below we describe in detail the substantial 
   modifications to the code. 

   \subsection{Variable smoothing length}

      The basic idea of the variable smoothing length method 
      \citep{Price04,Springel02} is that the smoothing length $h$ is 
      related to the particle coordinates through a relation between $h$ 
      and the particle density $\rho$
      \begin{equation}
         \frac {\partial h_a} {\partial \vec{r_b}}   = 
          \frac {\partial h_a} {\partial \rho_a}
          \frac {\partial \rho_a} {\partial \vec{r_b}}.
      \end{equation}
      The ansatz on the dependence of $h$ on $\rho$ is of the form
      \begin{equation}\label{hansatz}
         h_a = \eta \left ( \frac {m_a} {\rho_a} \right )^{1/3},
      \end{equation}
      where $\eta$ is a dimensionless parameter that specifies the
      size of the smoothing length in terms of averaged
      particle spacing (setting $\eta$ to $1.2$ gives in 3D about
      $60$ particles around any given one). The derivative of the above 
      equation respect to $\rho$ gives
      \begin{equation}
         \frac {\partial h_a} {\partial \rho_a} = 
          - \frac {h_a} { 3 \rho_a }.
      \end{equation}
      The density definition in VINE is given by
      \begin{equation}\label{rhodef}
         \rho(\vec r_a) = \sum_{b=1}^N m_b W(\vec r_{ab},h_{ab}),
      \end{equation}
      where $r_{ab}=|\vec r_a - \vec r_b|$ and $h_{ab}=(h_a+h_b)/2$, so we 
      have a nonlinear equation to be solved for both $h$ and $\rho$.\\
      To find a self--consistent solution to 
      eq.~(\ref{hansatz}--\ref{rhodef}), we have to solve the following 
      equation
      \begin{equation}
         f(h_a) = \rho_a(h_a) - \rho_{sum}(h_a) = 0,
      \end{equation}
      with a Newton--Raphson method until convergence is reached.
      The derivative of the function $f(h_a)$ is
      \begin{equation}
         f'(h_a) = \frac {\partial \rho_a} {\partial h_a} -
          \sum_b m_b \frac {\partial W_{ab}} 
                 {\partial h} = - \frac {3 \rho_a} {h_a} \Omega_a,
      \end{equation}
      where
      \begin{equation}
         \Omega_a = \left [ 1 - \frac{\partial h_a } {\partial \rho_a} 
          \sum_b m_b \frac {\partial W_{ab}}{\partial h} \right],
      \end{equation}
      accounts for the gradient of the smoothing length.
      Convergence is assumed to occur when 
      $|h_{new} - h | / h_0 < \epsilon = 10^{-3}$. Due to the dependence 
      of $h$ on $\rho$, the equations of motion are changed accordingly
      \begin{equation}
         \frac {d \vec v_a} {dt} = 
          - \sum_b m_b \left( \frac {P_a}{\rho_a^2 \Omega_a} + 
          \frac {P_b}{\rho_b^2 \Omega_b}  + \Pi_{ab} \right)
          \grad_b W_{ab},
      \end{equation}
      where $\Pi_{ab}$ is the artificial viscous pressure term. 
      The continuity equation becomes
      \begin{equation}
         \frac {d \rho_a} {dt} = 
          \frac{1} {\Omega_a} \sum_b m_b (\vec v_a - \vec v_b)
         \cdot \grad_b W_{ab},
      \end{equation}
      and the energy equation
      \begin{equation}
         \label{lagrange4}
         \frac {d u_a} {dt} = \frac {P_a} {\Omega_a \rho_a^2} \sum_b m_b 
          \vec v_{ab} \cdot \grad_b W_{ab} + \frac{1}{2} \sum_b m_b 
          \Pi_{ab} \vec v_{ab} \cdot \grad_b W_{ab},
      \end{equation}
      where $\vec v_{ab}=\vec v_a - \vec v_b$.\\
      The artificial viscosity term is added a) to 
      correctly model shock waves that inject
      entropy into the flow over distances that are much shorter than
      a smoothing length and b) to simulate the evolution of 
      viscous disks.
      The $\Pi$ term broadens the shock        
      over a small number of smoothing lengths and correctly resolves
      it ensuring at the same time that the Rankine--Hugoniot equations are
      satisfied. In this way it prevents 
      discontinuities in 
      entropy, pressure, density, and velocity fields.
      SPH simulations \citep{Monaghan83} include both a linear 
      term (bulk viscosity), which dissipates kinetic energy as particles 
      approach each other to reduce subsonic velocity oscillations 
      following a shock, and a quadratic term (von Neumann--Richtmyer 
      viscosity), which convert kinetic energy to thermal energy 
      preventing particle interpenetration in shocks
      \begin{equation}
         \Pi_{ab}=\left \{
          \begin{array}{l}
             (-\alpha_{\textrm{SPH}}c_{ab}\mu_{ab}+\beta_{\textrm{SPH}}
             \mu_{ab}^2)/ \rho_{ab} \;\; \; 
             \vec v_{ab}\cdot\vec r_{ab} \leq 0,\\
             0 \qquad \qquad \qquad \qquad \qquad \qquad \; 
              \vec v_{ab}\cdot\vec r_{ab} > 0,\\
          \end{array}
         \right.
      \end{equation}
      where all quantities are symmetrized. The term $\mu_{ab}$ 
      plays the role of the velocity divergence,
      \begin{equation}
         \mu_{ab}=\frac{h_{ab}\vec v_{ab} \cdot \vec r_{ab}}{\vec r_{ab}^2
         + \eta^2 h_{ab}^2} f_{ab},
      \end{equation} 
      with $\eta << 1$ to prevent singularities as particles
      approach, while $f_{ab}$ \citep{Balsara95} is introduced to
      avoid large entropy generation in pure shear flows and is defined as
      \begin{equation}
         f_a = \frac{|<\grad\cdot\vec v_a>|}{|<\grad\cdot\vec v_{a}>|+
                |<\grad \times \vec v_a>| + \eta'},
      \end{equation}
      where again $\eta' << 1$ is a factor used to prevent singularities, 
      and $\alpha_{\textrm{SPH}}$ and $\beta_{\textrm{SPH}}$ determine
      the strength of the artificial viscosity. In general they are set
      initially to $0.1$ and $0.2$,
      respectively, but they can vary
      during the simulation, keeping only their ratio fixed 
      \citep{Morris97,Rosswog00}.\\
      The shear viscosity contribution deriving from the linear and quadratic
      artificial viscosity terms can be compared to 
      the \cite{Shakura73} viscosity $\alpha_{SS}$ as in \cite{Meru12}:
      \begin{equation}\label{eqmeru}
         \alpha_{\textrm{SS}}=\alpha_{\textrm{SS,lin}}+
          \alpha_{\textrm{SS,quad}}=\frac{31}{525}
          \alpha_{\textrm{SPH}} \frac{h}{H} + \frac{9}{70\pi}
          \beta_{\textrm{SPH}}
          \left(\frac{h}{H}\right)^2.
      \end{equation}
      It has been shown \citep{Mona85,Lodapri10,Meru12} 
      that even in the continuum 
      limit, the artificial viscosity terms $\alpha_{\textrm{SPH}}$ and 
      $\beta_{\textrm{SPH}}$ mimic a Navier--Stokes viscosity.

   \subsection{Variable softening length}

      The variable softening length method is needed when 
      self--gravity is included in the model \citep{Price07}. 
      The modified gravitational potential per unit mass may be written 
      in the form
      \begin{equation}
         \Phi(\vec{r}) = -G \sum_b m_b \phi (|\vec{r} - \vec{r_b}|),
      \end{equation}
      where $\phi$ is a softening kernel that is a function of the
      particle separation and the softening length ($h$, which corresponds 
      to the smoothing length as in \cite{Price07}). The form
      of $\phi$ is given below.
      The gravitational force in $\vec{r}$ is computed as
      \begin{equation}
         \vec{F ( \vec{r} ) } = - \nabla \Phi = -G 
         \sum_b m_b \phi' (|\vec{r} - \vec{r_b}|)
         \frac{\vec{r} - \vec{r_b}} {|\vec{r} - \vec{r_b}|},
      \end{equation}
      where $\phi' = \partial \phi / \partial |\vec{r} - \vec{r_b}|$, and
      we have neglected the spatial variation of $h$.
      To get an expression for $\phi$, the Poisson's equation is used
      \begin{equation}
         \nabla^2 \Phi = 4 \pi G \rho,
      \end{equation}
      and by using the SPH definition for the density (eq.~\ref{rhodef}),
      it is possible to
      derive a relation between the smoothing kernel $W$ and
      the softening kernel $\phi$
      \begin{equation}
         W(r) = - \frac {1} {4 \pi r^2} 
          \frac{\partial} {\partial r} \left ( r^2
          \frac{\partial \phi } {\partial r} \right ).
      \end{equation}
      The kernel softens the gravity from neighbor particles
      while it is the usual $1/r$ potential for those far away.
      The additional terms to the equation of motion due to the 
      self--gravity are \citep{Price07}
      \begin{align}
        \frac{d \vec{v}_a}{dt} = &-G \sum_b m_b \phi'_{ab}(h_{ab}) 
         \frac{\vec{r_a}-\vec{r_b}}{|\vec{r_a}-\vec{r_b}|} \nonumber \\
         &- \frac{G}{2} 
         \sum_b m_b \left[\frac{\zeta_a}{\Omega_a}+
         \frac{\zeta_b}{\Omega_b}\right] W'_{ab}(h_{ab}),
      \end{align}
      where the first term represents the softened gravitational force, 
      and the second one is used when the adapting softening length is 
      adopted and restore the energy conservation with
      \begin{equation}
         \zeta_a=\frac{\partial h_a}{\partial \rho_a}\sum_bm_b
          \frac{\partial\phi_{ab}}{\partial h}.
      \end{equation}

   \subsection{SPH implementation of radiation hydrodynamics in the 
               flux--limited diffusion}

      The coupled energy equations describing the evolution of the 
      specific gas internal energy $u_g$ and of the total 
      frequency--integrated radiation energy $\xi$ are the following 
      \citep{Mihalas84,Whitehouse04}:
      \begin{align}
         \frac {D \xi} {Dt} & =
          - \frac {\grad \cdot \vec F} {\rho} - \frac { \grad \vec v : 
          {\vec P}} {\rho} - a c \kappa \left[ \frac {\rho \xi} {a} - 
          \left( \frac {u_g} {c_v} \right )^4 \right], \label{radeneqn} \\
         \frac {D u_g} {Dt}  & =
          - \frac {P \grad \cdot \vec v} {\rho} + a c \kappa \left[ 
          \frac{\rho \xi} {a} - \left( \frac {u_g} {c_v}
          \right )^4 \right], \label{spceneqn}
      \end{align}
      where $a = 4 \sigma_B / c$ and $\kappa$ is the mean absorption
      opacity. 
      The term $\frac {\grad \cdot \vec F} {\rho}$ on the right--hand side of 
      eq.~(\ref{radeneqn}) is the radiation flux term. For an isotropic 
      radiation field (\emph{Eddington approximation}), the radiative 
      flux is given by
      \begin{equation}\label{fluxdiffusion}
        \vec F = - \frac{c}{3 \chi \rho} \grad E,
      \end{equation}
      where $\chi$ is the total opacity, which is the sum of absorption 
      and scattering 
      terms; $E$ is the radiative energy density $E = \xi \cdot \rho$; and 
      $c$ is the light speed.
      In optically thin regions where $\chi \to 0$ the Eddington 
      approximation fails ($\vec F \to \infty$), since photons travel 
      freely and their free paths may exceed characteristic lengths of 
      the system, making the radiation field anisotropic. In this case 
      the flux--limited approximation is used \citep{Levermore81} and the 
      radiation flux can be written as a Fick's law of diffusion
      \begin{equation}\label{fluxlimited1}
         \vec F = -D\grad E,
      \end{equation}
      where $D=c\lambda/(\chi \rho)$ is a diffusion coefficient, and the 
      dimensionless function $\lambda=\lambda(E)$ is the flux--limiter 
      defined as
      \citep{Levermore81,Whitehouse04}
      \begin{equation}
         \lambda(R) = \frac{2+R}{6+3R+R^2}.
      \end{equation}
      where $R$ is the dimensionless quantity $R=|\grad E|/(\chi\rho E)$.
      In the optically thin limit ($R\to \infty$) the flux limiter tends 
      towards
      \begin{equation}
         \lim_{R\to\infty} \lambda(R) = \frac{1}{R},
      \end{equation}
      so the magnitude of the flux approaches 
      $|\vec F|=c|\grad E|/(\chi \rho R)=cE$.
      In the optically thick (or diffusion) limit, $R\to 0$ so the 
      flux--limiter
      \begin{equation}
         \lim_{R\to 0} \lambda(R) = \frac{1}{3}
      \end{equation}
      and the flux takes the value given by eq.~(\ref{fluxdiffusion}).\\

      The second term on the rigth--hand side of eq.~(\ref{radeneqn}) 
      $\frac { \grad \vec v : {\vec P}} {\rho}$, describing the radiation 
      pressure, contains the radiation pressure tensor $\vec P$ that, 
      in the flux--limited approximation, can be expressed in terms of 
      the radiation energy density
      \begin{equation}\label{fluxlimited2}
         \vec P = \vec f E,
      \end{equation}
      where $\vec f$ is the Eddington tensor, defined as
      \begin{equation}\label{fluxlimited3}
         \vec f = \frac{1}{2}(1-f)\vec I + \frac{1}{2}(3f-1)\hat{\vec{n}} 
          \hat{\vec{n}},
      \end{equation}
      where $\hat{\vec{n}} =\grad E/|\grad E|$ is the unit vector in the 
      direction of the radiation energy density gradient and $f=f(E)$ is 
      a scalar function called the Eddington factor. The flux limiter and 
      the Eddington factor are related through
      \begin{equation}\label{fluxlimited4}
         f = \lambda + \lambda^2 R^2,
      \end{equation}
      The last term in both eqs.~(\ref{radeneqn}) and (\ref{spceneqn}) 
      controls the interaction between the radiation and the gas. In fact 
      we can rewrite the term inside the square bracket as $T_r^4-T_g^4$.

      In non--irradiated protoplanetary disks, the temperatures 
      are low enough that the energy stored in radiation is negligible 
      compared to the thermal energy of the gas ($ \xi << u_g$). Under 
      this condition, the so--called one--temperature approach
      \citep{Kley09},  the  two coupled equations, eqs.~(\ref{radeneqn}) 
      and (\ref{spceneqn}), reduce to a single equation for the gas 
      internal energy $u_g$
      \begin{equation}\label{intenergy}
         \frac{Du_g}{Dt} = 
          - \frac {P \grad \cdot \vec v} {\rho} - 
            \frac {\grad \cdot \vec F} {\rho}.
      \end{equation}
      The radiation flux term can be rewritten using the flux--limiter 
      $\lambda$ as
      \begin{align}
         - \frac {\grad \cdot \vec F} {\rho} 
          & = \frac {1} {\rho} \grad \cdot \left(
              \frac{c \lambda}{\kappa_R \rho} \grad E \right), \\
          & = \frac {1} {\rho} \grad \cdot \left[
              \frac{c \lambda}{\kappa_R \rho} \grad \left(
              \frac{4\sigma_B}{c}T^4\right)\right], \nonumber \\ 
          & = \frac {1} {\rho} \grad \cdot \left[\frac{16 \sigma_B \lambda}
              {\kappa_R \rho} T^3 \grad T \right], \label{eneq}
      \end{align}
      where $\kappa_R$, the Rosseland mean opacity, approximates $\chi$.
      To compute $\kappa_R$ we use a power--law dependence on temperature 
      and density $ \kappa_R = \kappa_0 \rho^a T^b$ as in \cite{Bell94}
      where the coefficients $\kappa_0$, $a$, and $b$ depend on the opacity 
      regime.
      The previous equation has the same form of 
      the heat conduction equation
      \begin{equation}
         \frac{du}{dt} = \frac {1} {\rho} \grad \cdot (k \grad T),
         \label{heatcond}
      \end{equation}
      where $k$ is the thermal conductivity. This similarity is useful 
      when we try to implement the energy equation (eq.~\ref{eneq}) in the
      SPH formalism. \cite{Cleary99} give in fact the following SPH 
      expression for the heat transport equation 
      \begin{equation}
         \frac {du_a} {dt} = \sum_{b=1}^{N} \frac{m_b} {\rho_a \rho_b}
          \left( \frac {4 k_a k_b} {k_a + k_b} \right ) 
          (T_a - T_b) \frac { \grad W_{ab}} {r_{ab}}.
          \label{radflux}
      \end{equation}
      In the case of radiative diffusion, the coefficients $k_a$ and $k_b$
      have to be substituted by \citep{Whitehouse04}
      \begin{equation}
         k_a=\frac{16 \sigma_B}{\kappa_a \rho_a} \lambda_a T_a^3.
      \end{equation}

      \subsubsection{Cooling via boundary particles}

         While the disk is heated by the viscous dissipation
         (we neglect star irradiation effects), it is at the same 
         time cooled by the thermal emission into the vacuum at the upper 
         and lower surfaces. To model the radiation escape, we introduce 
         ``boundary'' particles into the algorithm. These particles 
         populate the regions where the optical depth of the disk $\tau$
         is equal to 1 \citep{Ayliffe09,Meru10}. To identify the 
         boundary particles among all the particles representing
         the disk we proceed as follows:
         \begin{enumerate}
            \item The disk is divided in sectors 
                  ($r + \Delta r$, $\theta + \Delta \theta$);
            \item In each sector, SPH particles are sorted along both 
                  $r$ and $z$; 
            \item The parameter $z_b$ is defined as the height 
                  at which the optical 
                  depth $\tau$ is $\approx 1$ so that
                  \begin{equation}
                      \tau = - \int_\infty^{z_b} \kappa \rho dz.
                  \end{equation}
                  Under the assumption of low temperature and density in 
                  the outer layers of the disk, the vertical isothermal 
                  approximation can be locally used (thin disk condition)
                  so that the opacity $\kappa$ is approximately constant 
                  and can be taken out of the integral leading to the 
                  following expression
                  \begin{equation}
                     \tau = \kappa \int^\infty_{z_b} \rho dz = \kappa 
                      \Sigma_b = 1,
                  \end{equation}
                  where $\Sigma_b$ is the surface mass density of the 
                  boundary particles;
            \item In the SPH formalism, the superficial density can be 
                  computed as
                  \begin{equation}
                    \Sigma = \frac{N m_p}{A_s},
                  \end{equation}
                  where $m_p$ is the SPH particle mass, $N$ the total 
                  number of particles present in the sector (independently 
                  of the $z$ coordinate), and $A_s$ the sector area;
            \item Knowing $\Sigma_b$ we can compute $N_b$,  the number of 
                  particles populating the layer with $\tau = 1$. This 
                  number is given by
                  \begin{equation}
                     N_b=\frac{A_s}{\kappa m_p},
                  \end{equation}
                  This is done in both the positive and negative vertical 
                  directions.
         \end{enumerate}

\begin{figure}
   \centering
   \includegraphics[width=0.5\textwidth]{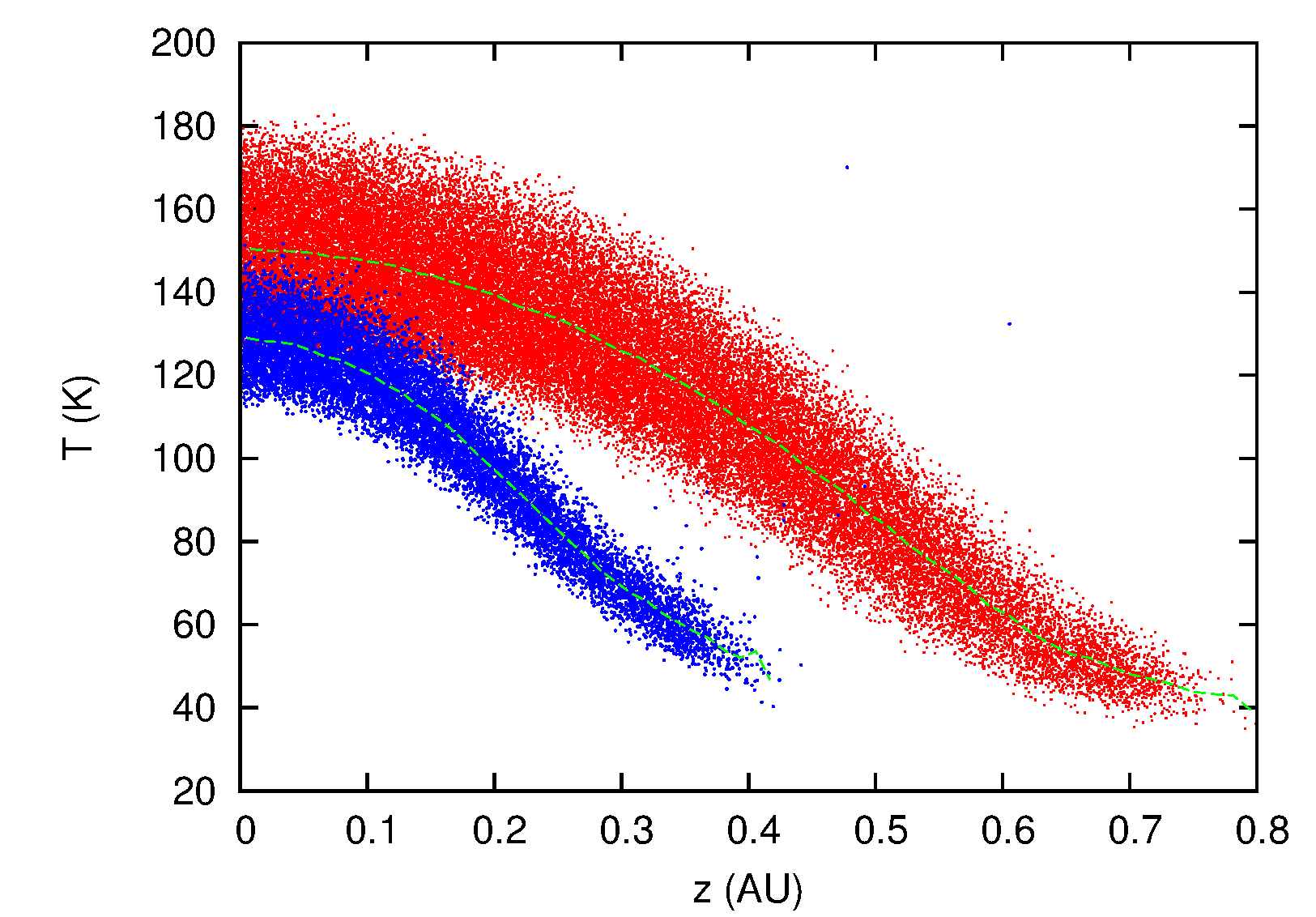}
   \caption{Temperature profiles as a function of the
   height over the median plane of the disk (z component)
   for the primary disk at $ r \sim 6$ AU and for the secondary
   disk at $r \sim 3$ AU. These locations are far from the
   center of the disk where the heating due to the spiral waves, even at
   apocenter, is the strongest. The dots are the individual
   temperatures of the SPH particles encompassed between
   $6$ and $6.3$ AU for the primary disk and $3$ and $3.3$ AU for
   the secondary disk. The continuous lines are the
   average temperature values. The SPH particles of the
   primary disk are represented by red dots, while blue
   dots are relative to the secondary disk.
}
   \label{fig0}
   \end{figure}

         The boundary particles evolve normally, but they interact radiatively 
         with the ``normal'' SPH particles in the bulk of the disk absorbing 
         radiation without releasing it. The amount of heat they 
         are supposed to radiatively transfer to their neighbors, which is
         computed from the energy equation, is lost to infinity. In this 
         way they act as cooling particles radiating away energy. It is 
         noteworthy that the temperature of the boundary particles is not 
         set to a low value but is in equilibrium with the local 
         temperature profile of the disk. 
         This prevents them from
         rapidly migrating towards the median plane of the disk due to their 
         low pressure value. However, they absorb the energy of the other 
         particles and act as energy sinks. 
         In Fig.~\ref{fig0} we show two representative vertical temperature
         profiles obtained from our simulations of disks in binaries. They
         are computed in a quiet ring far from the center of both the 
         primary and secondary disks to avoid the heating due to strong
         spiral waves, which also act when the binary is at the apocenter. 
         The cooling due to boundary particles is effective and
         leads to a decrease in the temperature towards the surfaces of 
         the disk. 


\section{Initial setup of the disks in the binary system}\label{setup}

   \begin{table}[tp] %
   \caption{Parameters of the simulations}
   \label{modeltab}
   \centering %
   \begin{tabular}{|c|c|c| }
   \hline
   Acronym & $a_B$ (AU) & $\rho$ (g/cm$^3$) \\
   \hline
   HIDECL  & 30         & $1 \cdot 10^{-9}$ \\
   \hline
   LODECL  & 30         & $5 \cdot 10^{-10}$ \\
   \hline
   HIDEFA  & 50         & $1 \cdot 10^{-9}$ \\
   \hline
   LODEFA  & 50         & $5 \cdot 10^{-10}$ \\
   \hline
   \end{tabular}
   \end{table}

   The parameter space of a binary system with a circumstellar disk 
   surrounding each stellar component 
   is very wide so we focus in this paper on a configuration that is supposed
   to be the more plausible one according to observations. A mass ratio 
   of $\mu = M_s/M_p = 0.4$ and an eccentricity of $e=0.4$ are adopted in 
   the definition of our standard case since these values are 
   statistically the
   most frequent among the binary systems observed so far \citep{dunma91}.
   Two different values of the binary semi-major axis are adopted in our 
   models: $a=30$ and $a=50$ AU. Coplanarity is assumed between the disks 
   and the binary orbit, even if in the future we plan to relax this 
   constraint. 
   We consider a high--density disk with 
   a midplane density 
   $\rho_0(r=1,z=0) = 1\cdot10^{-9} g/cm^3$, which is a value close to that 
   prescribed by the 
   minimum mass solar nebula model. A second less massive case,
   with $\rho_0(r=1,z=0) = 5\cdot10^{-10} g/cm^3$, is   
   also modeled to test the influence of the initial density profile on the 
   morphology and physical properties of the disks.
   Hereinafter, the four different runs are named in the following 
   way: the high--density 
   case with binary semi--major axis $a_B = 30$ AU is called HIDECL, 
   while the low density case 
   with the same value of $a_B$ is named LODECL. The two runs with 
   $a_B = 50$ AU are called HIDEFA and LODEFA, respectively.\\
   The density midplane 
   radial profile is computed as $\rho_0(r)\propto r^{-1.5}$.
   The initial vertical density stratification of the disks is
   computed as in \cite{Bitsch10}
   using a Gaussian--like dependence of the density on z. This 
   is a good approximation for a stationary disk where the pressure 
   balances the z--component of the central star gravity
   \begin{equation}
      \rho(r,z)=\rho_0(r,z) exp\left[-\frac{z^2}{2H^2}\right],
   \end{equation}
   with the scale height $h=H/r$ initially set to a constant value of 0.04. 
   The circumprimary disk extends from 
   $0.5$ AU to $8$ AU and the circumsecondary to $5.5$ AU, both within the 
   tidal truncation limits computed by \cite{Arty94} in the case where the 
   binary semi-major axis is set to $30$ AU. For the second case where 
   $a_B = 50$ AU 
   our configuration with relatively small disks is similar 
   to that observed for the system L1551 IRS 5 
   \citep{rod98} where the circumstellar disks of the binary system are 
   well separated and smaller than the tidal truncation radius. 
   We did not increase
   the size of the disks up to the tidal truncation radius since we wanted 
   to explore whether  a less perturbed configuration with the stars 
   moving farther
   away affects the same disks of the simulations with $a_B = 30$ AU.
   At the outer border, the disk density is smoothly 
   truncated with $\rho(r,z)$ exponentially decreasing.
   The mass of the primary disk in the HIDECL and HIDEFA models 
   is $0.015 M_{\odot}$, while that of the secondary is $0.003 M_{\odot}$.
   The secondary disk is less massive since we scale its initial density with 
   the stellar mass and its initial radius is also smaller. In the 
   LODECL and LODEFA models the masses of the disks are 
   $0.0075 M_{\odot}$ and $0.0015 M_{\odot}$, respectively since the 
   density is reduced by a factor 2.\\
   Each disk is initially 
   evolved as a single disk around its star until it reaches a steady state 
   both in density and temperature. Once this state is reached, the stars and 
   their disks are combined into a binary system with the desired orbital 
   parameters. Each SPH simulation uses about $1 \cdot 10^6$ particles 
   distributed in both disks according to their size and density. The 
   distinctive parameters of the four different models are summarized in 
   Table~\ref{modeltab}.
   Those SPH particles traveling within $0.5$ AU are accreted by the star. 
   Their number is limited in time, and for this reason, we do not 
   correct for their effects on the pressure and viscous forces 
   on particles moving inside the border as in \cite{batbopri95}.\\

   The initial values of
   $\alpha_{\textrm{SPH}}$ and $\beta_{\textrm{SPH}}$
   adopted in the simulations are 
   0.1 and 0.2, respectively. The  
   $\alpha_{SS}$ viscosity, estimated from 
   eq.~(\ref{eqmeru}), is about $(2-4) \cdot 10^{-3}$ 
   for the primary disk and $(1-2) \cdot 10^{-3}$ for the disk 
   around the secondary star. The $\alpha_{SS}$ of the primary disk is 
   very close to that of \cite{Nelson00}.


\section{Results}\label{results}

   In the forthcoming sections, we describe the evolution of the 
   disks in the four different models. We focus on the disk 
   morphology,the occurrence of shock waves, and the mass exchange between 
   the two disks (in particular for the 
   cases with $a_B = 30$ AU). 
   Shock waves  cause the onset of hydraulic jumps along the 
   vertical axis, which might 
   significantly affect the dust accumulation towards the median 
   plane of the disk and possibly inhibit 
   the planetesimal formation via pairwise accumulation. In addition, 
   in the models with small $a_B$, the high temperature induced by 
   the pericenter passages,
   and not
   fully dissipated at the apocenter, may further inhibit planet 
   formation, as already guessed by \cite{Nelson00}. The asymmetry 
   in the disk shape will also be explored as a potential prediction 
   for observers. 

   \subsection{High--density disks}
    
\begin{figure}
   \centering
   \includegraphics[width=0.5\textwidth]{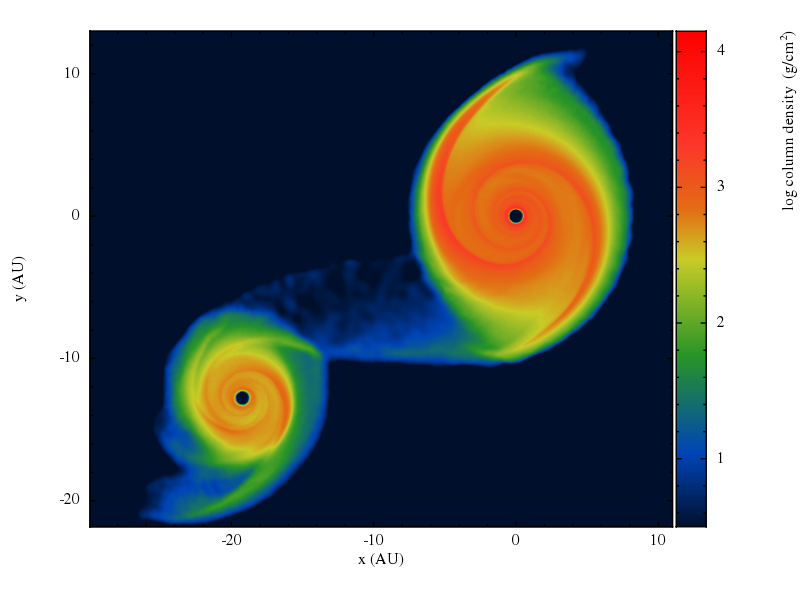}
   \includegraphics[width=0.45\textwidth]{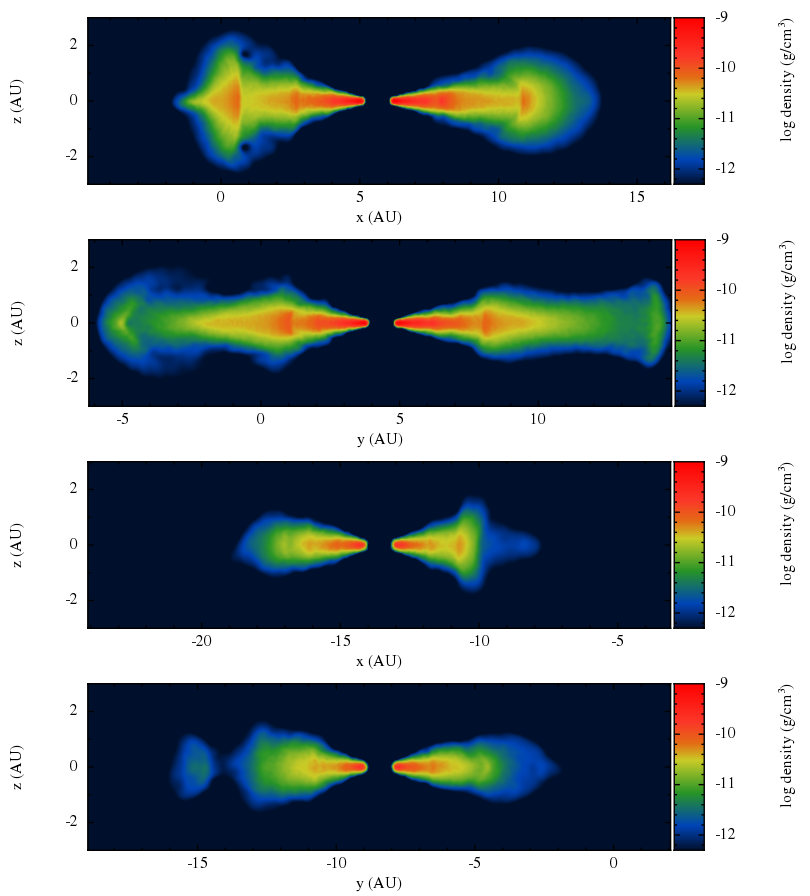}
   \caption{In the top plot the logarithm of the superficial density 
   (column integrated) of the 
   circumstellar disks for HIDECL (high density, small 
   binary semi-major axis) is shown during the 
   pericenter passage of the secondary star. In the bottom panels
   the density (non--integrated) is drawn along the $x-z$ and $y-z$ sections 
   of the primary disk (2nd and 3rd panels) and secondary disk 
   (4th and 5th panels).}
   \label{fig1}
   \end{figure}

   In the top panel of Fig.~\ref{fig1} we illustrate the integrated 
   density of the 
   circumstellar disks in the HIDECL model. The secondary star is 
   close to the pericenter of its orbit around the primary,  
   the most perturbing configuration for the two disks. 
   This is the third pericenter passage of our simulation, which
   took about three months of CPU time on a dedicated 32--processor machine 
   (VINE is parallelized with OPENMP). This long integration time is 
   due to the radiative transfer, which requires a very short time step,
   in particular close to the pericenter when strong
   spiral 
   waves are excited in both disks, and mass is transferred from one 
   disk to the other. 
  
   The primary disk displays two prominent, tidally generated 
   trailing spiral arms that
   tightly wind towards the center of the disk. The appearance of this
   pattern, initially at the outer edges of the disk, 
   begins when the companion star approaches the primary star,
   and they reach their maximum intensity shortly after the pericenter
   passage.  They are transient features, and they are gradually 
   damped when the stars evolve towards the apocenter. 
   The gas density is significantly higher at the location 
   of the spirals as is the temperature (see Fig.~\ref{fig5}). 
   The arms cover a substantial portion of the disk affecting its 
   overall evolution even by locally changing the shear viscosity 
   and heating rate. 
   In Fig.~\ref{fig1} the secondary disk seems to have a more 
   complex spiral pattern with three arms, but an additional arm, as we
   see later on, is simply excited by  the impact of 
   stream material coming from the 
   primary disk. A significant mass transfer occurs during the 
   pericenter passage mostly from the primary star to the secondary. 
   This phenomenon might in the long term lead to a redistribution of 
   mass between the two disks with the initially smaller one becoming slowly 
   more massive and gradually loosing memory of its initial density profile. 
   The spiral waves are also visible in the  
   $x-z$ and $y-z$ sections of the disks as regions of higher 
   density and inflated in the vertical direction.  

   \subsection{Spiral shock waves, hydraulic jumps and dust settling}

      Three--dimensional waves in accretion disks act like fundamental 
      modes that
      correspond to large surface distortions in the disk 
      \citep{Lubow98}.
      \cite{Boley05} have shown that shock waves in 3D 
      disks cause 
      abrupt increases in the disk scale height. These jumps 
      convert part of the flow's initial kinetic energy into 
      potential energy, while some is irreversibly lost 
      into heat. Breaking waves, generated at the 
      jump, crash  
      onto
      the pre--shock flow creating additional disordered motion and 
      possibly affecting the
      chondrule formation and dust settling processes.
      This shock--related splashing, observed in the bottom plots 
      of Fig.~\ref{fig1}, 
      is evidently highly nonlinear and has the 
      characteristics of hydraulic jumps that behave, in part, like 
      gravity modes \citep{Martos98}.
      
\begin{figure}
         \centering
         \includegraphics[width=0.4\textwidth]{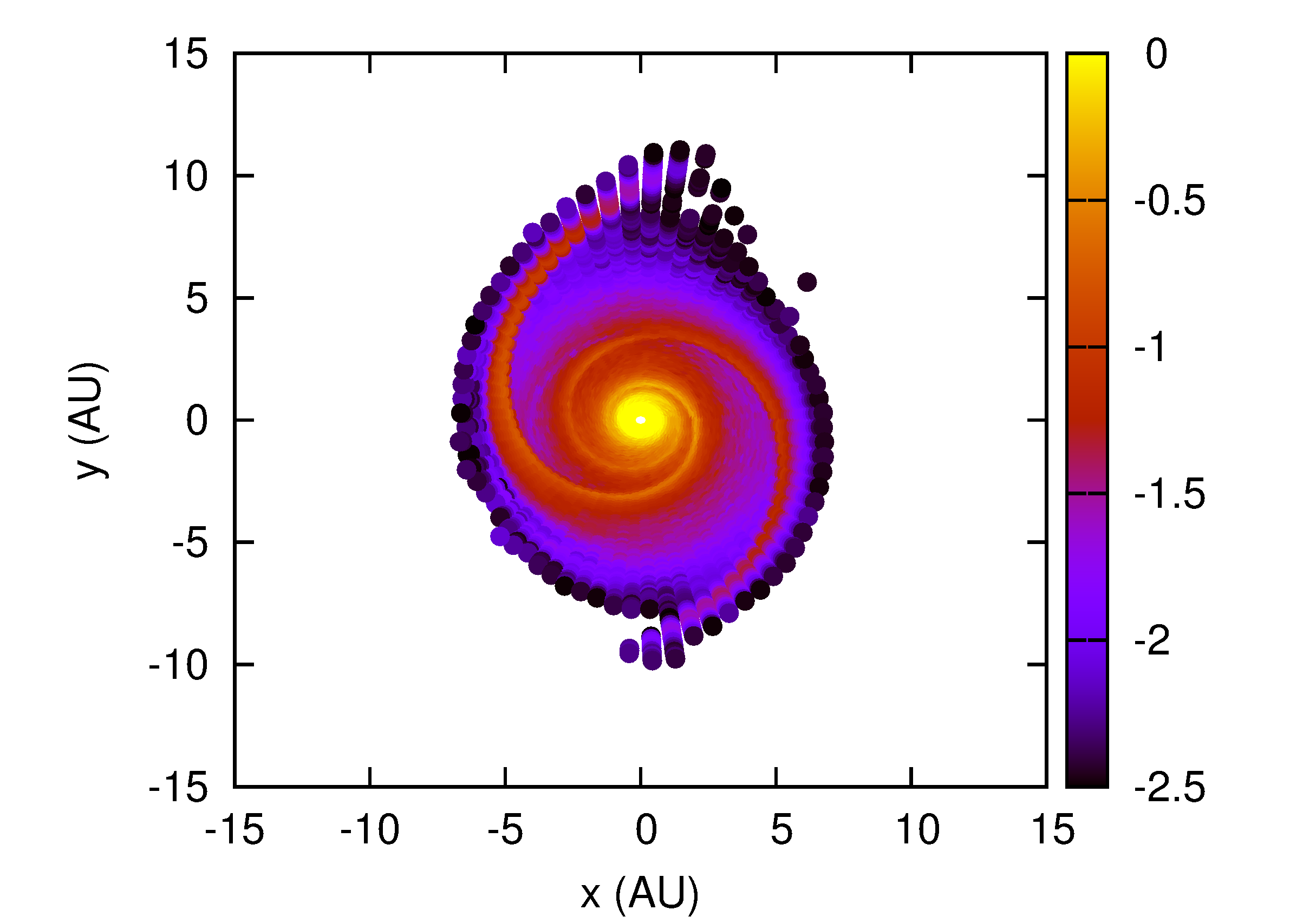}
         \caption{Vertically integrated vortensity in the primary disk 
                  during the pericenter passage in the HIDECL model.}
         \label{FigVort}
      \end{figure}

      To verify that the waves excited during the pericenter passage are
      indeed shock waves, we performed two tests. First we estimated 
      a 2D vortensity by computing the ratio between a 
      column integrated vorticity  $\omega$ derived for each SPH 
      particle as  $\omega = \hat{\vec{z}} \cdot 
      (\nabla\times\vec{v})$ and the superficial density $\Sigma$ as
      the average of the 3D density $\rho$ over concentric rings
      \begin{equation}
        \label{eq:vortensity}
        \eta=\frac{\omega}{\Sigma}.
      \end{equation}
      This quantity is used locally as an indicator of shocks since 
      vortensity is generated when the material passes through a shock.
      In Fig.~\ref{FigVort} the spiral waves are clearly outlined, and 
      the vortensity perturbations are superimposed on the density waves 
      observed in the superficial density plots. 
 
      \begin{figure}
         \centering
         \hskip -0.7 truecm
         \includegraphics[height=0.35\textwidth,width=0.49\textwidth]{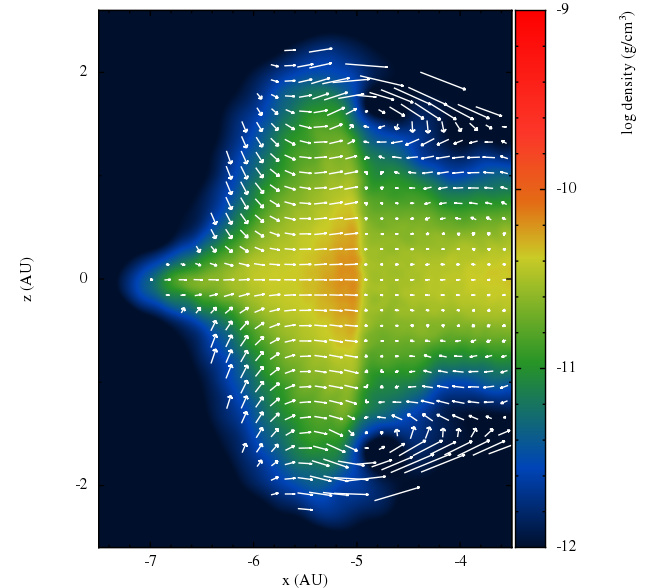}
         \hskip -0.5 truecm
         \includegraphics[height=0.35\textwidth,width=0.49\textwidth]{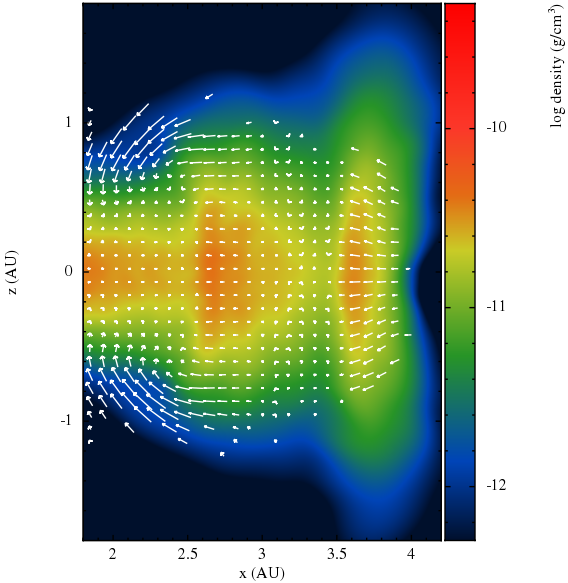}
         \caption{Velocity field and density of a slice in the $(x,z)$ 
         plane in the region of hydraulic jumps for the primary (top 
         panel) and secondary (bottom panel) in the HIDECL model. Both
         disks are centered on their parent stars to properly compute 
         the gas velocity field. The scales in $x$ and $z$ are 
         different in the two plots and,  in addition, the star is on the 
         right in the top plot (circumprimary disk), while it is on the 
         left in the bottom one (circumsecondary disk).}
         \label{fig4}
      \end{figure}

      The second test relies entirely on the fact that spiral shocks 
      also lead to hydraulic jumps \citep{Durisen11}. 
      In general, a hydraulic jump occurs when 
      the flow reaches a region 
      where there is an abrupt decrease in its speed. Conservation of 
      mass, momentum, and energy requires that the bulk kinetic energy of 
      the prejump flow be converted across the jump 
      into  
      gravitational potential energy and disordered motion.
      There is an increase in the flow vertical height at the 
      shock wave passage, 
      according to the Rankine--Hugoniot equations, and the 
      consequent fallback of gas onto the disk causes additional 
      disordered motion and the developing of vortices.
      Even though the gas in the protoplanetary disk is compressible, 
      the formation of a shock wave, in our scenario forced by the 
      gravitational perturbations of the companion star, produces a 
      similar phenomenon. We can estimate quantitatively 
      how much the scale height of the disk is affected by the shock wave using
      a simplified model developed by \cite{Boley05}, which assumes that the 
      shock is planar, is vertically stratified in the direction perpendicular 
      to the wave propagation, and has the pre--shock region in vertical 
      hydrostatic equilibrium.\\
      Defining the jump factor $J_f$ as the ratio between the pressure
      body forces and self--gravitating potentials in the pre--shock and
      post--shock regions \cite{Boley05} find that for strong shock waves 
      ($M>>1$) and when the 
      background potential dominates the gas potential
      \begin{equation}
        J_f\to\frac{2\gamma M^2(\gamma-1)}{(\gamma+1)^2},
      \end{equation}
      where $M$ is the Mach number and $\gamma$ is the adiabatic index.
      The two limiting cases are
      \begin{itemize} 
        \item $J_f>1$, the gas is overpressured, and it will expand 
          vertically;
        \item $J_f<1$, self--gravity cause the gas to compress.
      \end{itemize}
      A strong adiabatic shock ($M\to\infty$) disrupts vertical 
      hydrostatic equilibrium, because $J_f>1$ and the material expands 
      upward at the sound speed on a timescale of approximately 
      $H/c_s\approx\Omega^{-1}=t_{rot}/2\pi$.
      As shown in the bottom panels of Fig.~\ref{fig1}, at the spiral waves the 
      density distribution along the z--axis is puffed up, and this confirms
      that in our models we are in the  $J>1$ case where the gas 
      expands vertically. 
      To predict the height of the shock bore, \cite{Boley05} 
      adopt a classical hydraulic jump model, where $g_z = cost$. The jump
      of the fluid behind the shock is determined by the Froude number
      \begin{equation}\label{Fraude}
        F = \frac{V}{c} = \frac{u_1}{\sqrt{g_zh_1}},
      \end{equation}
      which is defined as the ratio of a characteristic velocity ($V$) to a 
      gravitational wave velocity ($c$) or as the ratio of a body's inertia
      to gravitational forces, and it is an analogous to the Mach number.
      In a non--dissipative jump,
      \begin{equation}
         \frac{h_2}{h_1}=\sqrt{\frac{1}{4}+2F^2}-\frac{1}{2}.
        \label{jf0}
      \end{equation}
      In the limit of a strong jump ($F>>1$) we would have $h_2/h_1\sim F$.
      Using this classical result as a model for understanding the maximum
      height, a shock bore reaches during the post--shock vertical expansion,
      \cite{Boley05} derive, for a non self--gravitating disk,
      \begin{equation}
        \frac{h_2}{h_1}\approx\sqrt{J_f}.
        \label{jf1}
      \end{equation}
      This ratio has a behavior similar to that of the Froude number described
      in eq.~(\ref{jf0}). When $M=1$, $h_2/h_1\sim1$, and $h_2/h_1\sim M$ when 
      $M>>1$. This does not mean that any fluid element close to the 
      shock wave shows a jump in the vertical direction, but it does imply that 
      the disk scale height will change. As an example, material near the 
      midplane 
      ($d\Phi/dz\to 0$, $J_f\to0$) will not be affected by a single shock wave 
      passage, while higher altitude gas will have the strongest response.

      In our HIDECL case, we can apply this model to the 
      spiral waves observed in Fig.~\ref{fig1} and estimate the 
      jump factor $J_f$ and the corresponding Froude number. 
      In Fig.~\ref{fig4} we show a detail of both disks in the 
      $(x-z)$ plane around the shock wave. In Fig.~\ref{fig4}a 
      (upper plot),
      we show a slice of the circumprimary disk where the 
      star is on the right. In Fig.~\ref{fig4}b (lower plot) we
      instead show a detail of the circumsecondary disk where the star is 
      on the left. Different scales are used in the plots to magnify the 
      density variations. The formation of hydraulic jumps at the 
      spiral waves is clearly visible with a significant increase of 
      the gas vertical height and the formation of breaking waves on top 
      of the jumps. The motion of the gas is evidenced by velocity 
      vectors which are superimposed to the 2--D density plot. There is
      a significant decrease in the velocity magnitude across the wave
      and an abrupt change in the gas density. A rough estimate of the 
      jump--factor $J_f$ from Fig.~\ref{fig4}a by using eq.~(\ref{jf1}) 
      gives a value of 
      $J_f \approx 3.2$  and a Froude number (from eq.~\ref{jf0}) 
      $F \approx 1.6$.\\
      Similar values are also found in the secondary 
      disk. The velocity field evidences the formation of breaking 
      fronts at the top of the hydraulic jump and the splashing of 
      material from the top of the jump. Shock bores not only 
      generate the vertical displacement of fluid elements illustrated 
      in Fig.~\ref{fig4}a,b, but they also drive gas to large radial 
      excursions from their circular orbits, causing large amounts of 
      wave energy to be transformed into kinetic energy 
      stirring and mixing the disk. This is also at the origin of the 
      disk heating. When the gas crosses the shock front, the shock 
      normal component of the fluid element velocity diminishes, 
      according to the Rankine--Hugoniot equations, while the 
      tangential component is preserved and the flow become supersonic 
      after the shock. This leads to streaming along the spiral
      arms \citep{robe79}. Moreover, when the gas expands upwards, the
      pressure confinement normal to the shock loosens and the fluid 
      expands radially, causing some gas to flow back over the top of 
      the shock. The resulting morphology is a spiral pattern
      moving through the disk in the $(x-y)$, as illustrated in 
      Fig.~\ref{fig1}, while the pattern 
      appears as a breaking wave in the vertical direction. 

      In the inner part of the disk we do not observe breaking waves, 
      but this is due to the fast crossing of winding spiral arms. 
      The orbital period of a fluid element is
      much shorter than the pattern period of a spiral wave. Shortly 
      after the first shock, the gas therefore encounters another arm 
      before it can
      settle back onto the disk, ending up elevated between shock passages.
      However, in the outer part of the disk the periods become comparable
      and the shock bores have the time to develop into breaking waves.

      The evolution of the gas disk into hydraulic jumps may have critical 
      consequences for the dust settling towards the disk midplane. When the 
      gas is pumped up at the shock waves, via aerodynamic effects,
      it can drag the smaller components
      of the dust inverting their settling motion. As a consequence, 
      at each pericenter passage, the disk will develop strong 
      spiral waves able to stir up the dust significantly slowing 
      the sedimentation process down, if not suppressing it. In addition, 
      it would also increase the relative velocity between dust 
      particles halting the 
      coagulation process at small dust sizes and possibly preventing
      the formation of planetesimals through the conventional scenario
      of dust coagulation. On the other hand, turbulent motion that
      may develop in the proximity of spiral waves, may favor the fast 
      accretion of pebbles into large planetesimals, as suggested 
      by \cite{johan07} and \cite{cuzzi08}.

   \subsection{Temperature profile: Chondrule formation at shocks?}

\begin{figure}
         \centering
         \includegraphics[width=0.5\textwidth]{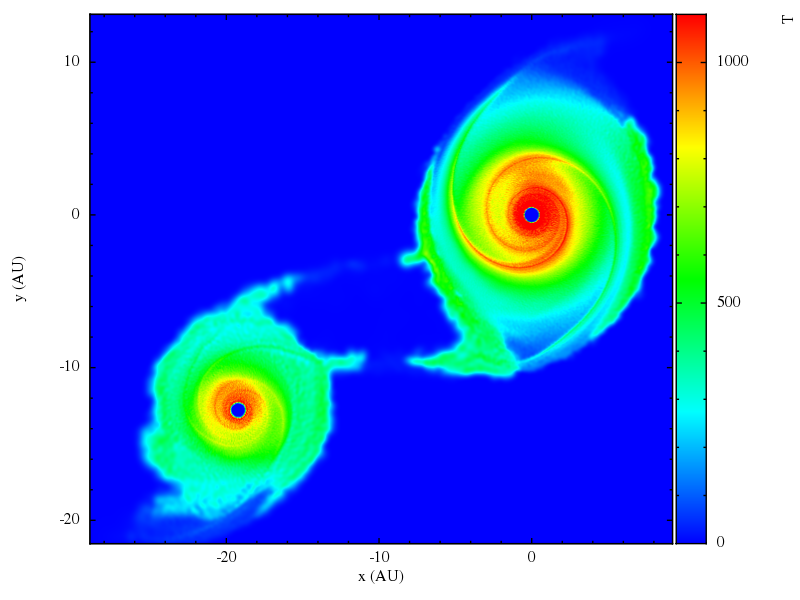}
         \includegraphics[width=0.5\textwidth]{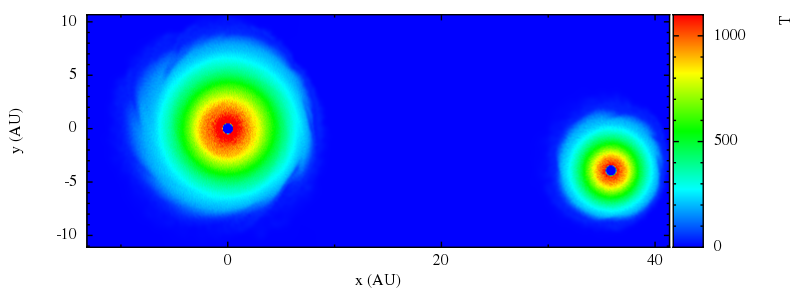}
         \caption{Temperature distribution in the two disks computed at 
         their median plane when the stars are close to the pericenter
         (upper panel) and apocenter (lower panel) in the HIDECL model.}
         \label{fig5}
      \end{figure}

      In Fig.~\ref{fig5} we show the midplane 
      temperature distribution of the two disks during the 
      pericenter passage. At the shocks generated by the spiral 
      structures, the temperature is raised to high values 
      that might cause vaporization of some grains, as
      suggested by \cite{Nelson00}, and chondrule formation
      \citep{Boley05}. 
      The secondary disk is cooler than the 
      primary and this is due to its lower density.
      It appears also overheated at the outer
      edge, more than the circumprimary, since  
      mass coming from the more massive circumprimary disk 
      strikes its outer borders thereby increasing its temperature.  

\begin{figure}
         \centering
         \includegraphics[width=0.5\textwidth]{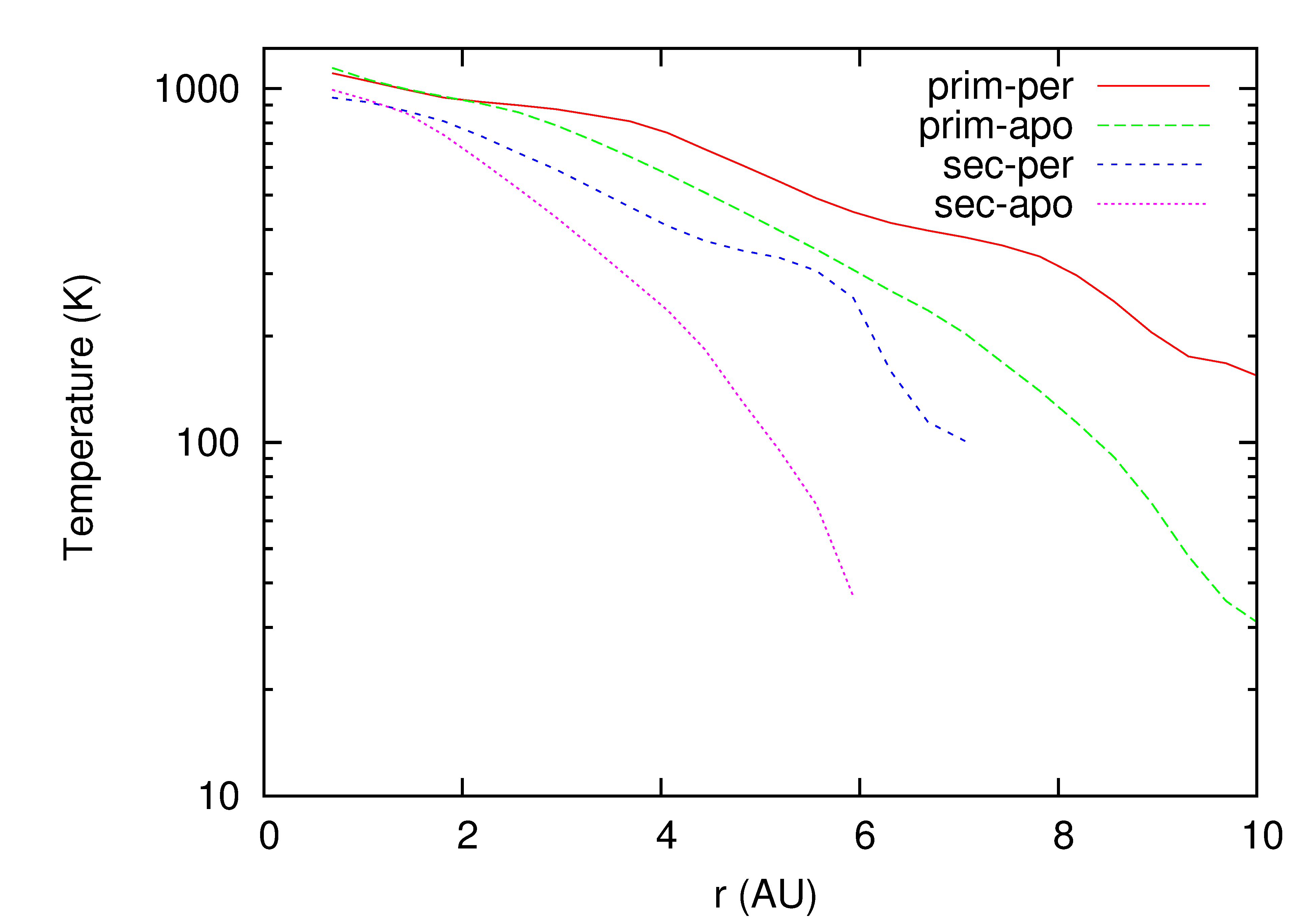}
         \caption{Azimuthally and vertically averaged temperature profiles
         of both disks in the HIDECL model
         when the stars are at pericenter and apocenter, respectively. 
         The secondary disk is less dense than around the primary,
         so its temperature is on average lower.
         }
         \label{fig6}
      \end{figure}

      During the pericenter approach, there is a considerable 
      transient internal thermal 
      energy generation in the disks  
      by means of shock waves and mass transfer. Once the stars depart 
      from each other traveling 
      towards the  
      apocenter,  the disks cool down due to radiative 
      cooling possibly reaching an equilibrium state. 
      This effect is shown in Fig.~\ref{fig6} where we 
      compare the azimuthally and vertically averaged temperature 
      profiles for both disks when the stars are at pericenter and 
      apocenter, respectively.  
      The difference is more marked in the outer parts
      of the disks, and it can be as large as 200 K. This phenomenon 
      was also 
      observed by \cite{Nelson00} in his simulations of an 
      equal mass binary system 
      with $a_B = 50$ AU and $e_b =0.3$. He performed 
      2D SPH numerical simulations of such a binary star/disk + star/disk 
      system finding a relatively mild difference in 
      temperature between pericenter and apocenter. Our larger 
      difference may be attributed to the different 
      dynamical configuration.
      In effect, our HIDECL model
      system is more compact 
      (smaller semimajor 
      axis), and it also has higher eccentricity. This 
      dynamical configuration leads to 
      a stronger heating at shock 
      waves and a larger mass exchange that
      causes a consistent local temperature increase 
      where the transferred flow impacts the disk. 
      Both these effects can explain the larger difference in 
      the temperature profiles between pericenter and apocenter 
      we find in our model. 
      On the other hand, we have lower temperatures in the disks 
      on average compared to the ones obtained in \cite{Nelson00} 
      but 
      this may be related to the different initial density adopted 
      in his simulations, the different orbital architecture and 
      the different cooling algorithm.
      The \cite{Nelson00} case compares better with our models where
      the stars have a larger separation ($a_B = 50$ AU), which
      is discussed later on. 
      Our temperature profiles appear to be comparable or slightly
      higher than those in \cite{muller12}, even if
      a different initial
      density profile is adopted 
      ($\Sigma \propto r^{-1}$ while ours 
      declines as $\Sigma \propto r^{-1/2}$). 
      In addition, \cite{muller12} consider
      a single disk around the primary,
      and as a consequence, the
      mass exchange between the two disks, 
      with its consequent heating, is not included in their model. 
      Also, since our simulations are
      performed in 3D, the amount of heating due to gas compression
      at the shock waves where  hydraulic jumps occur is 
      higher. 

      While high temperatures appear to prevent the condensation 
      of icy dust particles and then the growth of giant planet cores,  
      they may favor the formation of chondrules. 
      They form as melt droplets that were heated to high temperatures,
      while they were independent, free--floating objects in the protoplanetary
      nebula. After they were heated, cooled, and crystallized, chondrules
      were incorporated into the parent bodies in which chondrites originate.
      There are constraints on chondrule formation like a peak temperature 
      of about 1300 K followed by a fast cooling (100--1000 K per hour) as
      suggested in \cite{armi10}. In our simulations we find that 
      the temperature along the shock waves is higher than 1000 K, and it 
      drops quickly after the wave passage. This may favor chondrules 
      formation in such disks, in particular in more compact and eccentric 
      binary configurations where stronger and possibly hotter
      spiral waves develop. 
      This mechanism is not local since  
      the shock waves induced by the companion star covers the whole 
      disk, since the spiral wave extends from inside to the outer 
      borders. As a consequence, it would not be necessary to 
      invoke additional 
      heating mechanisms  or local shock fronts of different 
      origins \citep{
      urey53,urey67,
      sanders05,asphaug11,levy89,
      joung04,morfill93,pilipp98,desch00,
      nakamoto05,
      boss93,
      ruzma94,
      hood09,
      hood98, 
      weiden98,ciesla04,morris12,
      wood96,desch02,boss05,boley08}.
      In this picture, the strong shock waves generated by binary interaction
      during pericenter passages might be an additional and 
      very efficient mechanism for
      chondrule formation over the whole disk.

   \subsection{Mass exchange between the disks}

   When the two stars are at the pericenter, the gravitational 
   interaction of the companion on each disk is the strongest. 
   Spiral shock waves tidally induced by the gravitational perturbations 
   of the stars propagate within each disk causing a significant mass
   transfer between the two disks in addition to 
   disk heating. The mass exchange is 
   bidirectional, but in absolute value, the primary disk 
   donates more mass to the secondary, possibly because of its higher mass
   and its stronger shock waves that extend farther out from the disk center.

   Where the mass stream coming from one disk hits the other and is accreted,
   heat is generated. This effect can been seen in the temperature 
   distribution illustrated in Fig.~\ref{fig5}a.  At the edge of both disks, 
   where the exchanged mass is accreted, the temperature is locally 
   higher. This heating is subsequently spread around within the 
   disk, contributing to the overall disk temperature profile. 
   For this reason, in modeling disks in close binaries, it is important 
   to include the secondary disk since the final temperature 
   profile will depend on the amount of mass exchanged  
   between the disks.

\begin{figure}
         \centering
         \includegraphics[width=0.5\textwidth]{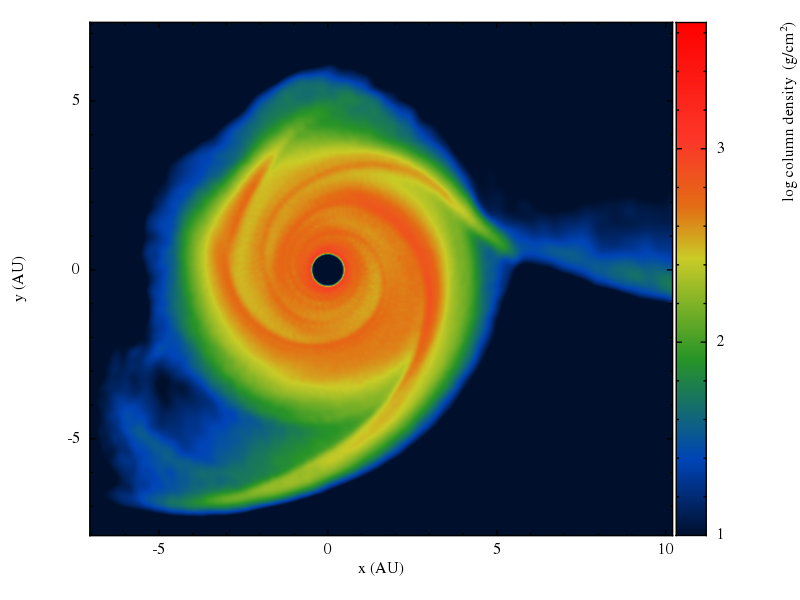}
         \caption{Density distribution of the secondary disk 
          when the stars are at the pericenter in the HIDECL model. 
          A new spiral arm is created by the material transferred 
          from the primary disk and impacting the outer edge of the 
          secondary.}
         \label{fig7}
      \end{figure}

   The mass accreted by the secondary disk generates an additional 
   spiral arm in the disk as illustrated in Fig.~\ref{fig7}. There 
   are three spiral arms in the disk, two generated by the primary star
   tidal field and an additional one produced by the the impact 
   of a stream from the primary. 

\begin{figure}
         \centering
         \includegraphics[width=0.5\textwidth]{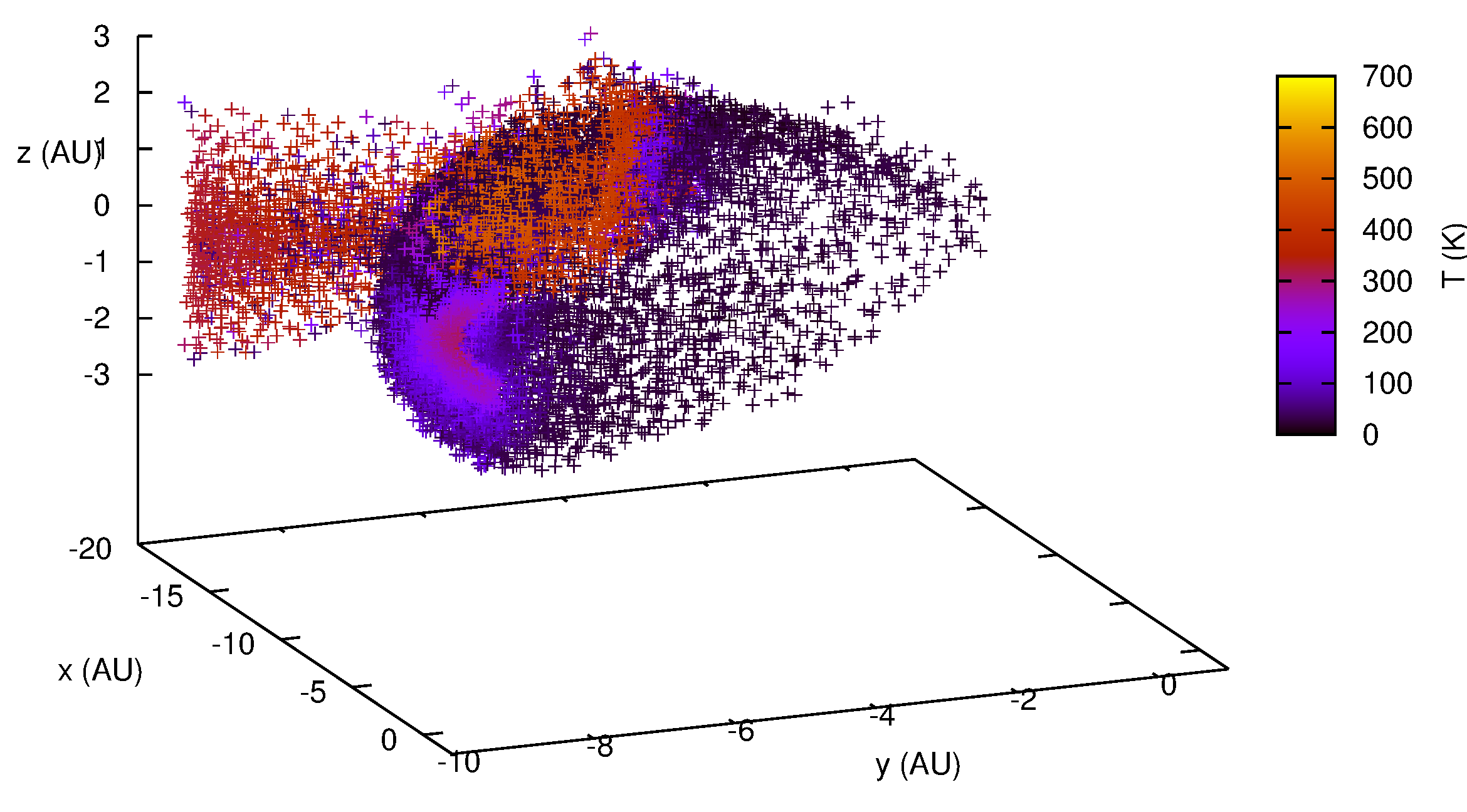}
         \caption{Temperature distribution of the 
          material flowing from the circumprimary disk 
          to the circumsecondary in the HIDECL model. The impact on the 
          secondary of the flow from the primary is 
          marked by a raise in the temperature. 
          }
         \label{fig7bis}
      \end{figure}

   The morphology of the mass flux moving from the primary 
   to the secondary reflects the shape of the shock wave. It 
   appears as a concave structure continuing in the shock 
   wave direction and propagating towards the secondary disk. 
   In Fig.~\ref{fig7bis} we show a 3D plot of the material 
   transferred from the primary disk to the secondary 
   and its temperature. The concave structure (like a water wave) is 
   visible, and the gas impacting the secondary disk is 
   heated up by the compressional motion. 

      The mass flow from the primary to the secondary disk 
      induced by the formation of spiral waves during the 
      pericenter passage can have significant
      effects in the long term. It can reduce the  
      viscous mass loss of the secondary disk and change
      the initial post--formation mass ratio of the 
      two disks. If this ratio was similar to the stellar 
      mass ratio during the protostar contraction, later 
      on this ratio could be different.
      The lifetime  
      of the secondary disk may be longer, thereby
      increasing the probability of finding a planet around a less 
      massive companion star, 
      as suggested by the discovery of the planet around 
      Alpha Centauri B \citep{dum12}. However, a longer 
      integration timespan is required to confirm this trend.

      \subsection{Disk eccentricity}

\begin{figure}
         \centering
         \includegraphics[width=0.5\textwidth]{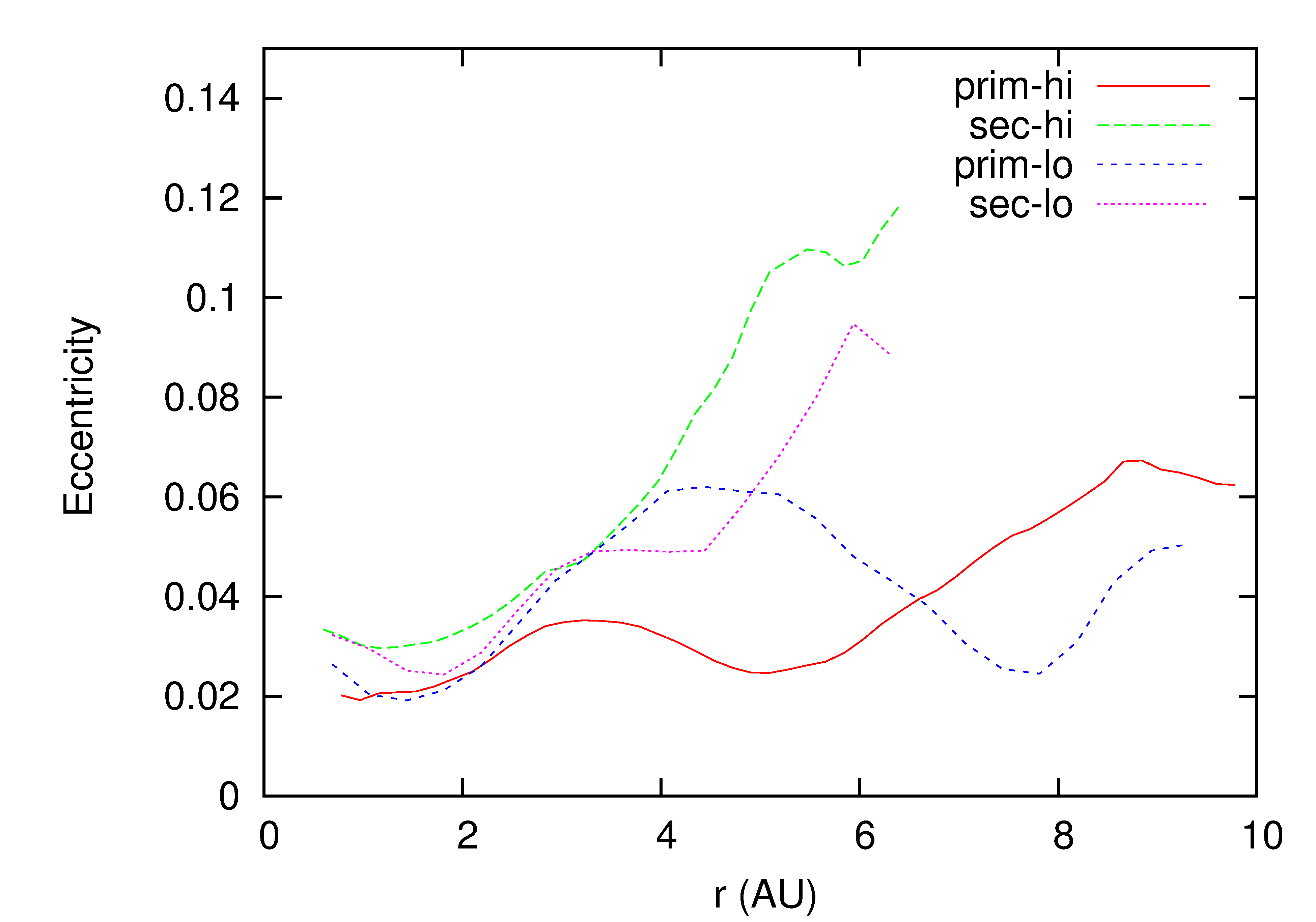}
         \caption{Disk eccentricity profiles, averaged 
         on azimuth and vertical direction, of the primary
         and secondary disks in the HIDECL and LODECL 
         models. The eccentricity is computed at the 
         apocenter when the strong spiral arms have almost
         completely dissipated.
          }
         \label{fig7tris}
      \end{figure}

      The disk eccentricity is an important parameter for 
      the evolution of a putative planetesimal population 
      born in the disk. As shown in \cite{paard08}, \cite{mabash2}, and 
      \cite{Kley08}, an eccentric shape for the disk may 
      perturb the evolution of the planetesimal eccentricity 
      and orbital alignment, which may lead to destructive collisions 
      rather than growth. The azimuthally 
      averaged eccentricity profile
      of both disks (primary and secondary) at apocenter, 
      where the spiral waves are dissipated, is shown in 
      Fig.~\ref{fig7tris} for the HIDECL model. The disk 
      eccentricity is low in the
      inner parts of both disks and it increases in the outer
      more perturbed regions. The eccentricity values agree
      with those derived in both \cite{muller12} and \cite{mabash2}
      for radiative disks on a longer timescale.
      In our model the secondary disk is significantly more eccentric 
      compared
      to the primary and this might be due to the stronger 
      perturbations of the primary star, which is more massive, 
      and to a reduced 
      self--gravity related damping effect
      \cite{mabash2}.  It is noteworthy that the 
      eccentricity profile we obtain after three pericenter passages 
      is similar to that of \cite{muller12} and \cite{mabash2}, 
      derived after more binary periods, making us confident that our
      results also hold in the long term.  
      In particular, 
      Fig.~3 of \cite{muller12} shows that, for radiative disks, 
      the eccentricity profile does not vary significantly with 
      time. This behavior is different from isothermal disks 
      where the disk eccentricity requires more time  
      to reach a quasi--stationary (and more 
      eccentric) state \citep{muller12,mabash2,Marzari09,Kley08}.
      As 
      suggested in 
       \cite{mabash2} and \cite{Cassen96}, 
      the energy loss
      by radiation may be at the origin of this 
      different evolution, leading to a 
      faster damping of density
      waves in radiative disks.

      \subsection{Low--density disks}

\begin{figure}
         \centering
         \includegraphics[width=0.5\textwidth]{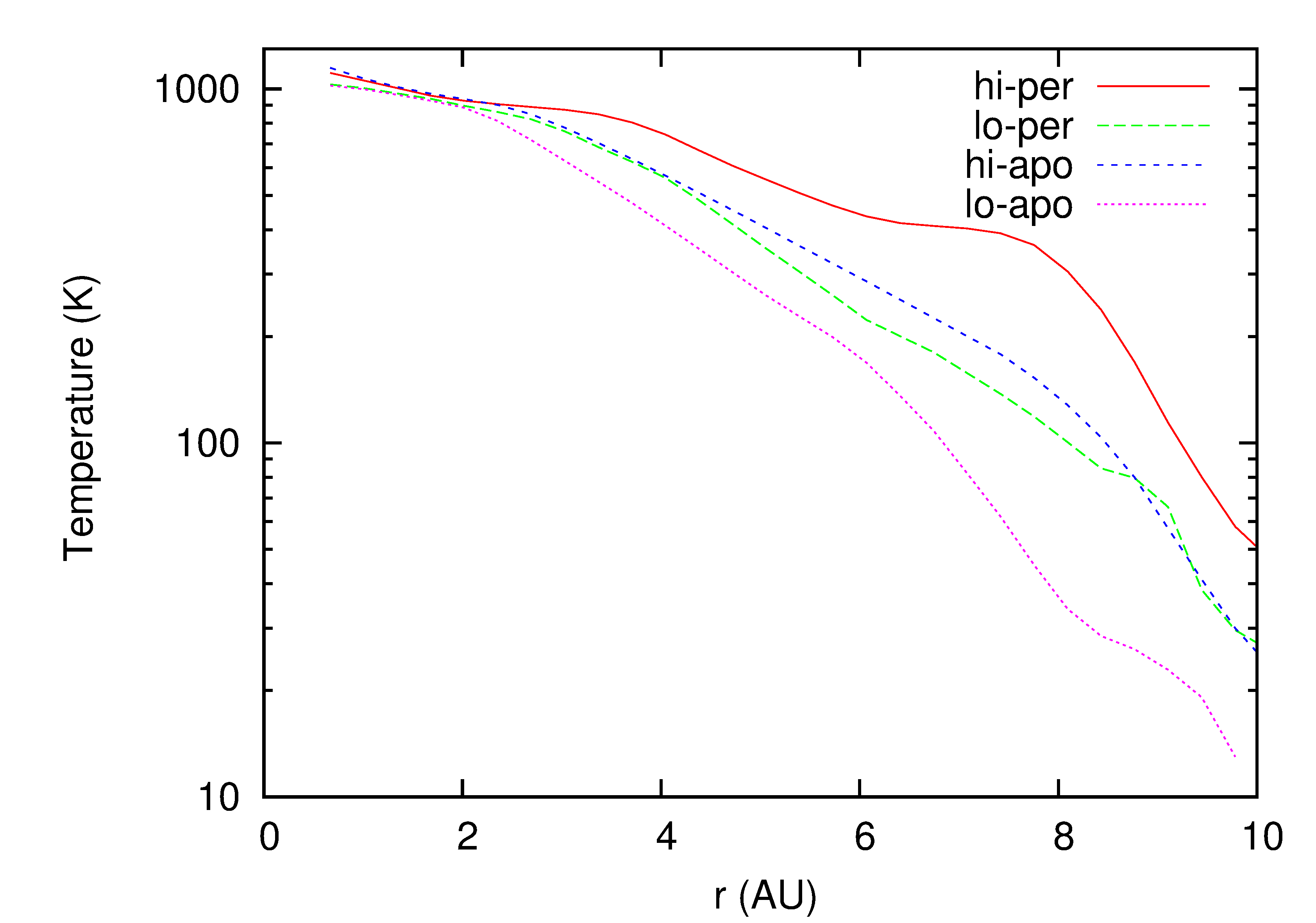}
         \caption{Azimuthally and vertically averaged 
         temperature profiles of the low--density disk (LODECL model)
         around the primary star
         compared to the high--density case (HIDECL) with the binary 
         at pericenter 
         and apocenter. The lower density disk is cooler in both 
         cases. 
         }
         \label{fig8}
      \end{figure}

\begin{figure}
         \centering
         \includegraphics[width=0.5\textwidth]{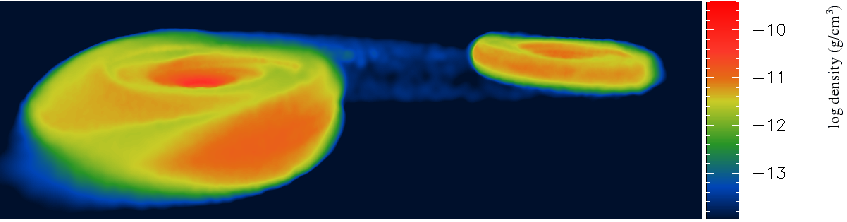}
         \caption{3D picture of the density distribution 
         of the disks in the LODECL model close to 
         the binary pericenter. Large vertical 
         jumps can be observed at the shock waves.  
         The surfaces are drawn at $\tau \sim 2$.
         }
         \label{fig9}
      \end{figure}

      The morphology of the disks in the LODECL model does not differ 
      significantly from that of the HIDECL and strong 
      trailing spiral
      patterns form in the both disks close to pericenter 
      dissipating by the time the stars move to the apocenter. 
      The main difference between the LODECL and HIDECL scenarios
      concern 
      the temperature distribution and the vertical height of 
      the hydraulic jump at the shock waves.  
      In Fig.~\ref{fig8} we compare the temperature profiles
      of the two cases with $a_B = 30$ AU. As expected, the 
      lower density case has an overall lower temperature, and 
      it is similarly heated up at the pericenter when shock 
      waves develop and mass transfer occurs. 
      Concerning the hydraulic jump at the shock waves, values 
      as high as 2.2 for the $h_2/h_1$ ratio are found in the 
      shock waves giving $J_f \sim 4.8$ and $F \sim 1.9$. These 
      values are slightly higher compared to the HIDECL case 
      and might be related to the lower sound speed in the 
      LODECL disk, leading to larger Mach numbers.  
      Figure~\ref{fig9} is a 3D picture of the two lower 
      density disks showing the large vertical jumps at the 
      shock waves. 

      The eccentricity of both disks is compared in 
      Fig.~\ref{fig7tris} to that of the HIDECL case 
      and they show a similar behavior. As a consequence, 
      the reduction of a factor two in density is not 
      enough to significantly affect the shape of the disks.
      Even if the self--gravity damping effect is less 
      strong \cite{Marzari09}, the lower temperature profile 
      of the LODECL case might 
      somehow help in keeping the overall disk eccentricity 
      \citep{mabash2} within 0.1 for both disks. 

      \subsection{Binary systems with larger separation}

      In the models HIDEFA and LODEFA the semimajor axis 
      of the binary system is increased to $50$ AU, a configuration 
      similar to the one explored by \cite{Nelson00}. The orbital 
      elements of the binary are the same, but \cite{Nelson00} 
      considers two equal mass solar type stars and equal mass
      disks, while we model a system where the secondary star 
      is less massive ($0.4 M_{\odot}$), and the disk is scaled 
      in density of the same ratio and is less radially extended. 
      Differences are then expected 
      in terms of physical properties of the disks in the simulations. 
      We did not match the configuration 
      of \cite{Nelson00} since we wanted to test the dependence 
      of the disk evolution on the binary semimajor axis so we kept the 
      same architecture of the previous models but we increase
      the binary semimajor axis. 

\begin{figure}
   \centering
   \includegraphics[width=0.5\textwidth]{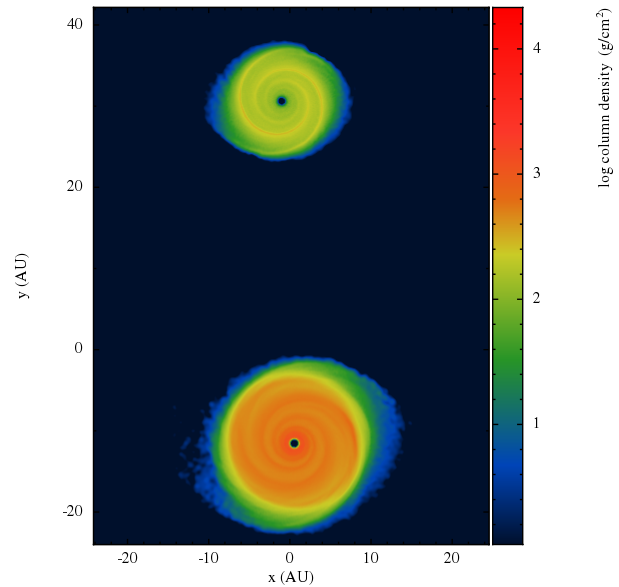}
   \includegraphics[width=0.45\textwidth]{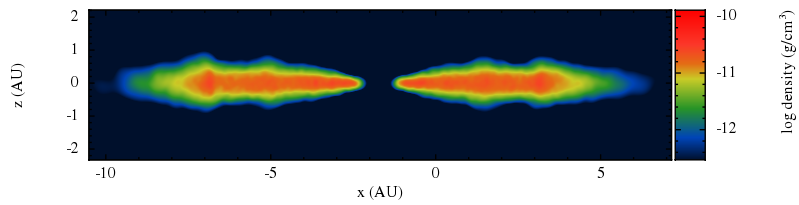}
   \caption{Integrated density profile of the disks in 
   the HIDEFA model shortly after the pericenter (upper
   panel). The non--integrated density of the disk around 
   the secondary star is shown in the 
   x--z plane. Shock bores are visible even if less 
   marked than in the HIDECL case. The hydraulic jumps 
   in the primary are of negligible height. 
   }
   \label{fig10}
   \end{figure}

      Figure~\ref{fig10} illustrates the integrated density distribution 
      of the two disks in the proximity of the binary pericenter in the
      HIDEFA model. Two--armed spiral shock waves are still 
      present in the secondary disk, while they are
      reduced to density waves in the primary. 
      These waves
      are weaker than those observed in the model 
      disks of 
      \cite{Nelson00}, owing to the lower mass of the companion star in our 
      simulations and to the different
      disk densities and sizes. The disk around the secondary shows
      shock bores as in the
      cases with $a_B = 30$ AU (HIDECL and LODECL).   This 
      is illustrated in the lower panel of Fig.~\ref{fig10} from 
      which a value of the jump factor $J_f \sim 2$ is estimated. 
      This value is significantly lower than that computed for 
      the HIDECL and LODECL cases as a consequence of the 
      less perturbative configuration in the HIDEFA model. 
      An almost negligible mass transfer between the two 
      disks occurs in this configuration, due to the lack 
      of strong shock waves on the primary disk. 

\begin{figure}
   \centering
   \includegraphics[width=0.5\textwidth]{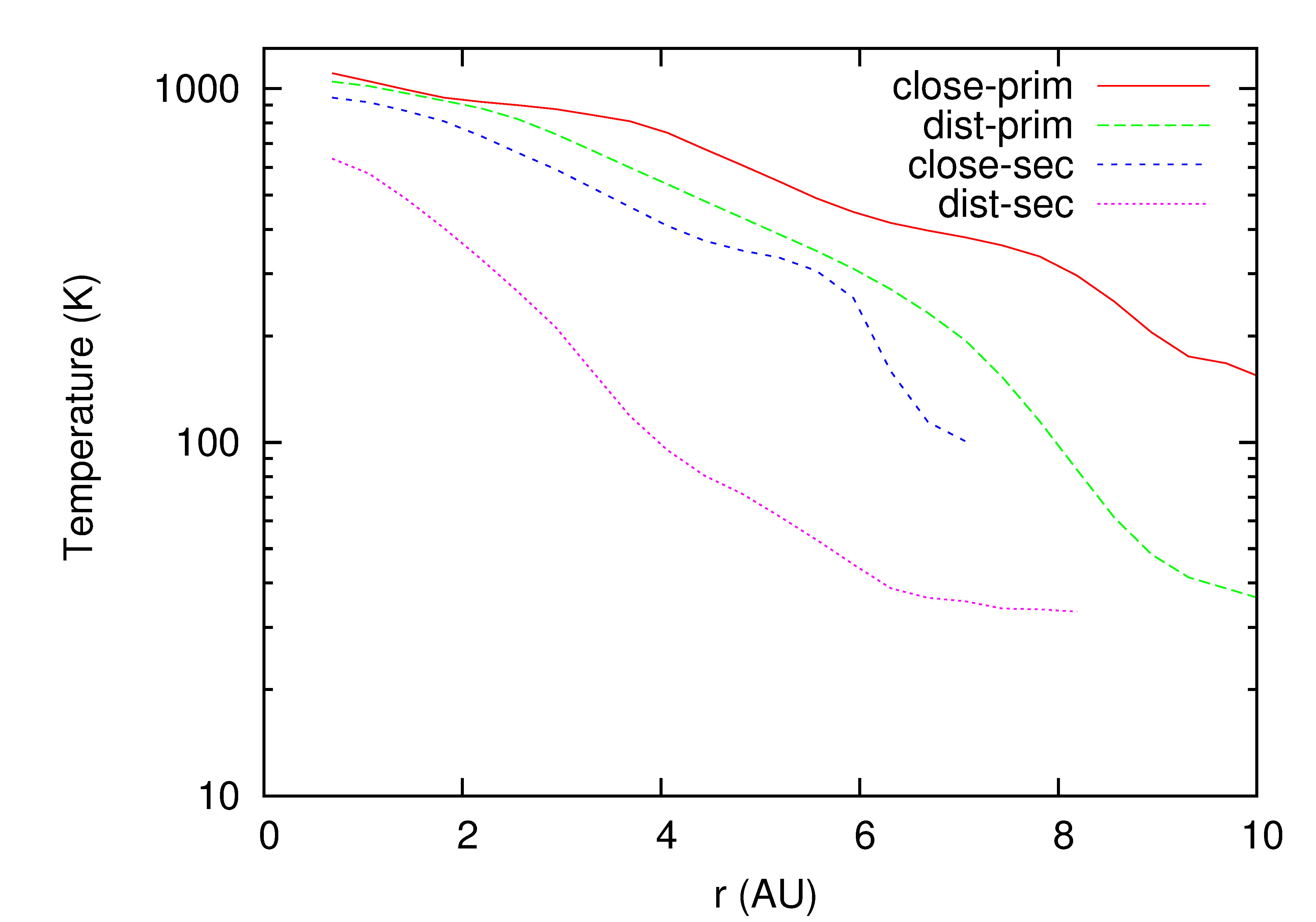}
   \includegraphics[width=0.5\textwidth]{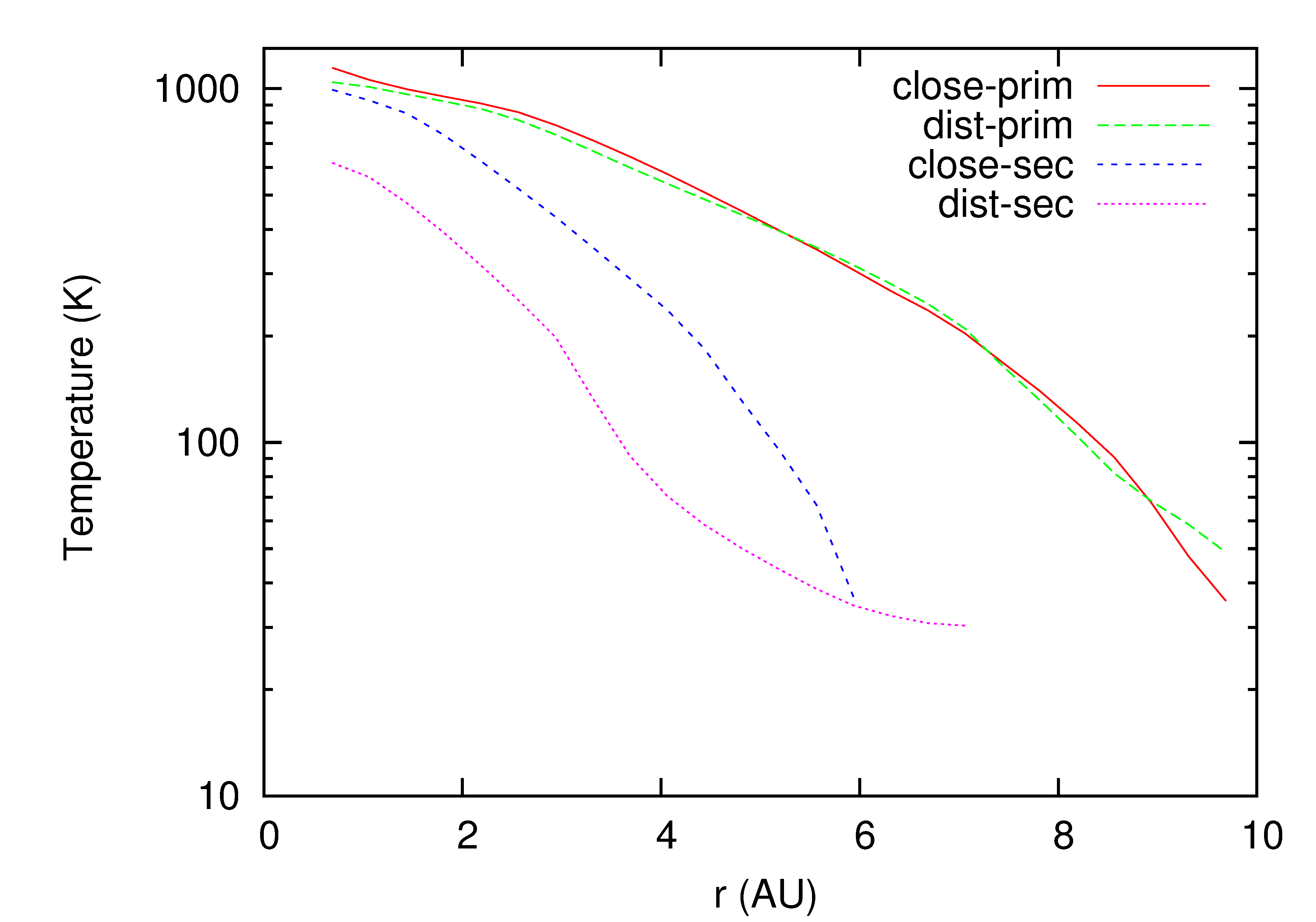}
   \caption{Comparison of the azimuthally and vertically 
   averaged temperature profiles in the HIDEFA and
   HIDECL cases at pericenter (upper panel) and 
   apocenter (lower panel) for both disks.  At apocenter 
   the temperature of the primary disk in the HIDEFA and
   HIDECL cases is very similar. 
   }
   \label{fig11}
   \end{figure}

      In Fig.~\ref{fig11} the averaged temperature profiles are 
      compared in the close and distant cases with the stars
      at pericenter and apocenter. At pericenter the close 
      case ($a_B = 30$ AU) is hotter compared to the 
      distant ($a_B = 50$ AU) case. This is an expected outcome 
      since both disks in the HIDECL case are significantly more 
      perturbed by shock waves, and a remarkable mass transfer occurs. 
      At apocenter, the primary disks have approximately the same 
      temperature profiles since the shock waves have dissipated, 
      and the viscous heating and cooling are almost in equilibrium. 
      The secondary disk is instead still very hot in the HIDECL 
      case, possibly because its dissipation timescale is longer than
      the binary orbital period, and its excitation at pericenter 
      was much stronger 
      than in the HIDEFA case, as also shown by the lower 
      value of $J_f$. Compared to the temperature profiles given in 
      \cite{Nelson00}, our values are lower, and this may be 
      ascribed to the differences in the architecture of the system,
      in the cooling algorithm and in the fact that our models 
      are 3D. 

\begin{figure}
         \centering
         \includegraphics[width=0.5\textwidth]{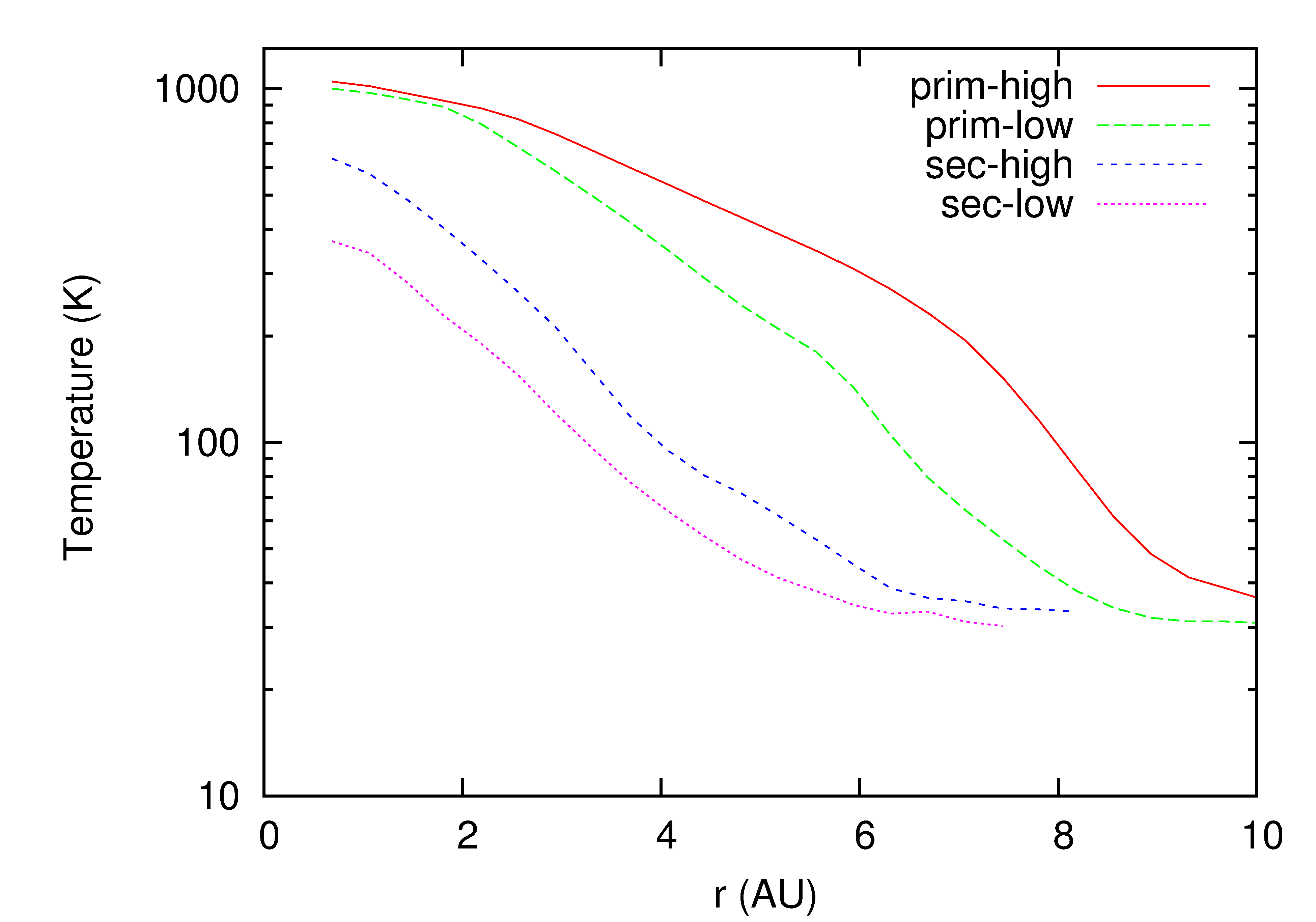}
         \caption{Temperature profiles at the binary pericenter 
         in the LODEFA and HIDEFA cases for both 
         the primary and secondary disks. 
         }
         \label{fig12}
      \end{figure}

      In Fig.~\ref{fig12} we compare the temperatures in  
      the LODEFA and HIDEFA cases at pericenter and, as expected, the 
      low--density disks are cooler. In this less perturbed 
      case, the difference in temperature between the 
      LODEFA and HIDEFA models must  
      be mostly ascribed to a different balance between viscous 
      heating and radiative cooling. In both cases the temperature 
      rise at pericenter is negligible when compared to the 
      close cases ($a_B = 30$ AU). This can be inferred by comparing
      the upper panel of Fig.~\ref{fig11}, where the stars are at pericenter,
      with the bottom panel where the stars are at apocenter. While there
      is a significant difference between the temperature profiles 
      of the close case (HIDECL) in the two dynamical configurations, 
      for the HIDEFA case the increase in temperature at pericenter 
      is significantly less marked for both disks.  
      

      \begin{figure}
         \centering
         \includegraphics[width=0.5\textwidth]{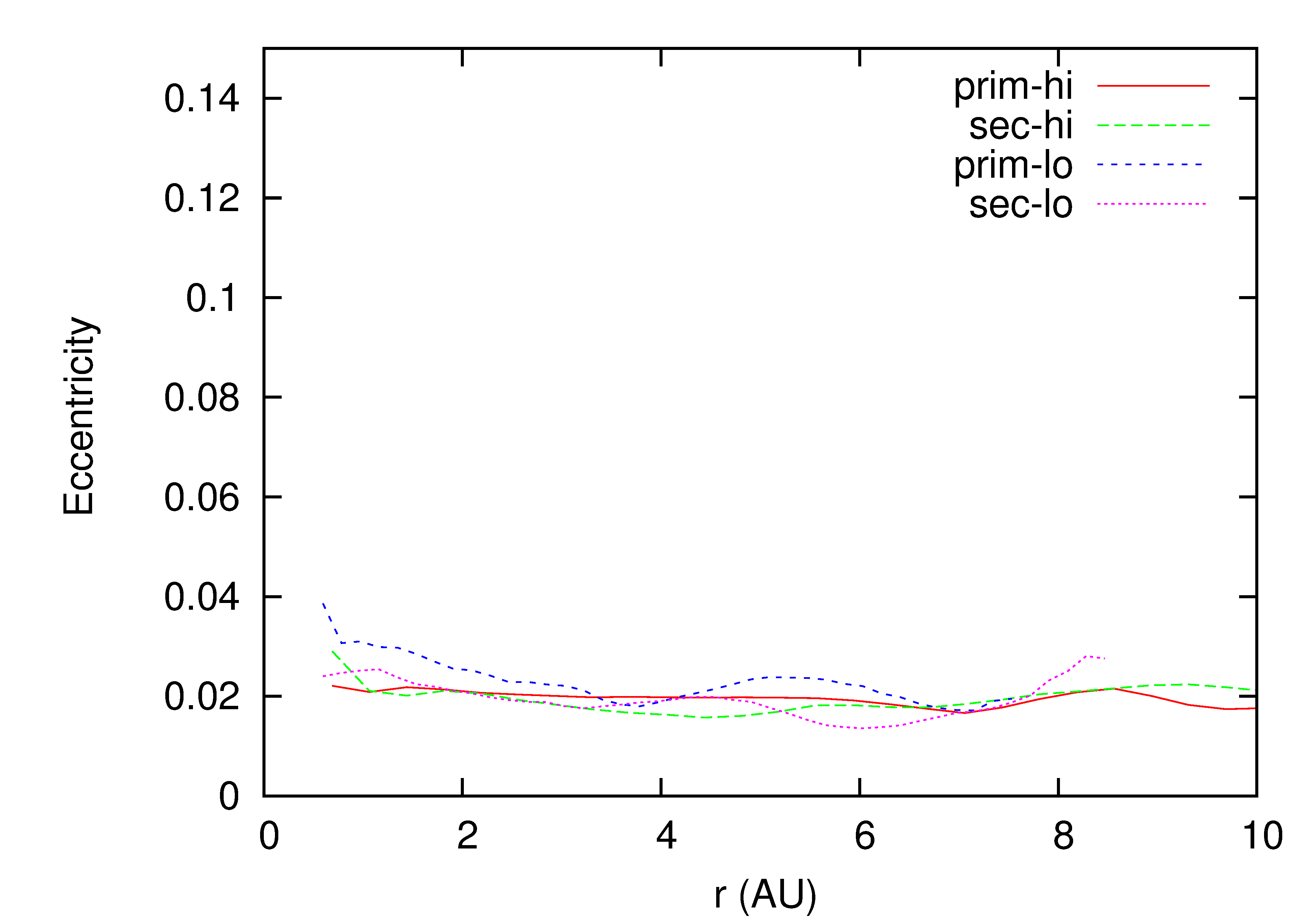}
         \caption{Eccentricity profiles for both 
         disks in the two distant
         ($a_B=50$ AU) HIDEFA and LODEFA cases. 
         The same scale of Fig.~\ref{fig7tris}
         is adopted for comparison. 
         }
         \label{fig13}
      \end{figure}

      In the distant configurations, the disk eccentricity 
      is small independently of the initial gas density, as 
      shown in Fig.~\ref{fig13}. We expect that in this more
      quiet configuration, planetesimal accumulation can proceed 
      less perturbed by the disk gravity.


\section{Summary and discussion}\label{conclusions}

We have performed 3D simulations of circumstellar disks
in binary systems using VINE, an SPH algorithm that has been 
modified to model a fully radiative disk. The cooling has been 
simulated by including boundary particles that populate 
the outer surfaces of the disks (defined by $\tau \geq 1$) 
and effectively radiate away the heat transported from the 
inner regions via radiative transfer. Four different 
binary and disk configurations are modeled to explore the 
influence of the companion star perturbations on the disk 
morphology, temperature, and eccentricity. Two close 
configurations ($a_B = 30$ AU) with different initial 
gas densities scaled by a factor 2 and two distant 
configurations ($a_B = 50$ AU) with the same scale 
factor in density. 

In the close configurations the spiral shock waves 
excited during the pericenter approach of the two stars 
generate hydraulic jumps whose height can be calculated from 
the simulations leading to an estimate of the Froude number. 
A significant mass transfer occurs between the disks at the 
binary pericenter when tidal trailing spiral waves are excited.  
This is an additional source of heat for both disks, and it  
can also increase the mass and lifetime of 
the smaller secondary disk enhancing the possibility of 
planet formation in it. It is also noteworthy that the 
mass ratio between the two disks at later stages of 
evolution may not reflect the primordial value due to 
the flux of mass from the primary to the secondary. 

In the 
secondary disks, the impact of material coming from the 
primary disks excites the formation of an additional shock 
wave. 
The temperature profiles 
show a large difference between pericenter and apocenter 
related to the heating generated by shock waves and mass transfer at 
pericenter. 
The disk eccentricity is relatively low for the 
primary disks with values compatible with previous 2D 
studies. The secondary disk is more eccentric compared to the 
primary in particular in the outer regions where the 
eccentricity can be as high as 0.1. 
There is no notable
difference in the disk morphology and disk eccentricity when 
we model less massive disks.
On the other hand, the temperature profiles, due to the 
dependence of the viscous heating on the density, are instead
significantly different. 

In the models with a higher value of the binary separation 
($a_B = 50$ AU), hydraulic jumps appear  
only in the secondary disks, while in the primaries the spiral 
waves, even at the pericenter, do not cause a significant 
change in the disk morphology. This implies that there 
is a limiting binary separation beyond which the
strength of the tidal spiral patterns on the primary disk
is not enough to excite the motion of the gas in the vertical 
direction and lead to mass transfer. 
In these less perturbed 
configurations, the disk eccentricity is lower even for the 
secondary disks with an upper limit of $\approx 0.03$ over
the full radial extent of the disks. Owing to the reduced strength 
of the spiral waves and negligible mass exchange, the 
distant configurations have lower temperature profiles 
with the stars at pericenter compared to the close
configurations. At apocenter the primary disks
in the distant and close configurations have the same temperature
while the secondary disk is hotter possibly because it did 
not dissipate all the heat accumulated during the pericenter
passage. As for the close configurations, the temperature profiles
of the less dense disks is lower than in 
higher density cases. 

\subsection{Implications for planet formation}

As already pointed out by \cite{Nelson00}, the high temperature 
of the disks and the consequent shifting outside of the ``snow 
line'' may inhibit the condensation of ices reducing the amount
of available mass for accreting the core of giant planets. 
Although in our models the temperature profiles are lower than 
those found by \cite{Nelson00}, even if with a different 
system architecture, the temperature is still too high in the 
close configurations. In the dense close configuration 
(HIDECL) at pericenter, the ``snow
line'' is at about $8$ AU for the primary
and $6$ AU for the secondary 
disk.  In both cases, the ``snow
line'' location almost coincides with the outer borders
of the disks. At apocenter, the two values shrink to $6$ AU and $4$ AU, 
respectively. In the more distant configurations, the ``snow
line'' is at $6$ AU for the primary with  negligible differences
between pericenter and apocenter, while it 
shifts from $2$ to $3$ AU for the secondary disk. Even locally, at the
shock waves, the temperatures are high enough to prevent icy 
dust condensation, but on the other hand, they might favor the 
formation of chondrules. 

In addition to the high temperature of the disk, our simulations 
show that the spiral waves generate hydraulic jumps that 
would invert the dust settling to the disk midplane induced by the 
aerodynamic drag. This would significantly affect the 
settling velocity and timescale and, as a consequence, the 
coagulation process into larger particles. The relationship 
between the size and vertical height would be destroyed and 
a substantial remixing of the dust would occur at the shock
bores. An additional nasty effect for particle growth would 
be the increase in the mutual relative velocities at the shock,
in particular where breaking waves crash back onto the disk. 

The two effects described above seem to work against the 
coagulation of dust into large planetesimals, thereby preventing 
the formation of planets in binaries. Even the subsequent
planetesimal accretion process appears critical in binary 
systems as already pointed out by 
\cite{mascho},\cite{thebs04},\cite{theb06},\cite{thems08},\cite{Kley08},
\cite{thems09},\cite{Marzari09},\cite{xie09},\cite{xie10},\cite{paard08},
and \cite{mabash2}.
However, 
close binary systems hosting giant planets have been detected, such as
$\gamma$ Cephei \citep{camp88}, Gliese 86 \citep{que00}, 
HD41004 \citep{zuck04}, HD196885 \citep{corr08}, and
the small terrestrial planet in Alpha Centauri
\citep{dum12}. There are two possible scenarios for the formation 
of the present dynamical architectures of these systems. 
The first is that the binary system hosting the planet had 
a larger separation in the past compatible with the growth 
of planets according to the  
standard core--accretion model. The subsequent dynamical evolution of the 
stellar system -- either because it is part of an unstable triple stellar 
system or because it suffered a close encounter with a third (background) star
\citep{marba07a,marba07b,marbe12} -- led to a shrinking of the 
binary orbit without ejecting the planet(s) from the system. 
A second possibility is related to the new model for planetesimal 
formation based on 
the direct formation of large planetesimals
from the accumulation of small solid particles in 
turbulent structures of the
gaseous disk
\citep{johan07,cuzzi08}. 
The onset of disordered motions
is strongly favored in circumstellar disks in 
binaries due to the formation of strong spiral waves that not 
only affect the radial evolution of the disk but also influence
the vertical structure of the disk, as shown in our models. 
The quick formation of large planetesimals may bypass the 
crucial stage of dust coagulation and subsequent planetesimal 
accumulation, even if it would not solve the problem of the high
temperature able to prevent the condensation of icy dust grains. 

\subsection{Speculations on the long--term evolution}

\begin{figure}
   \centering
   \includegraphics[width=0.5\textwidth]{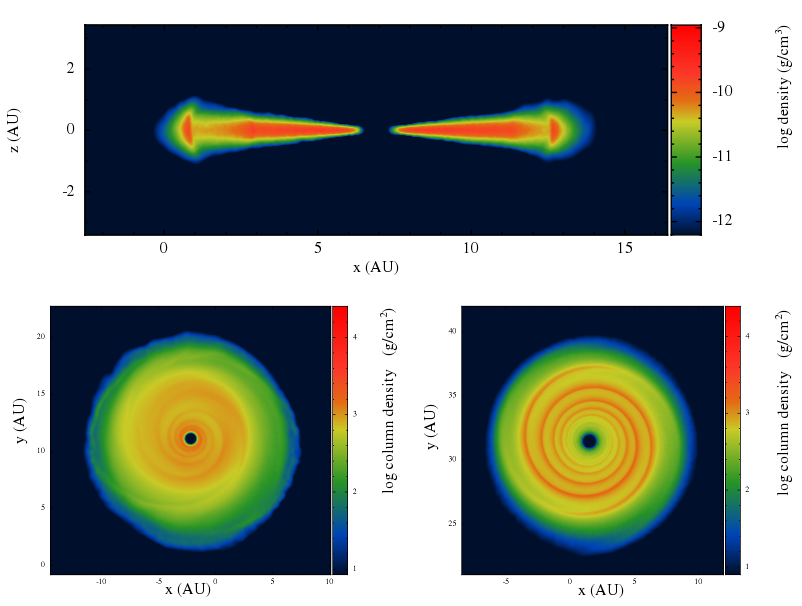}
   \caption{In the upper plot a vertical slice of an isothermal 
   disk with the HIDECL parameters is shown after 20 binary 
   revolutions near the pericenter passage. The two 
   lower plots, illustrating the primary disks shortly after 
   the binary pericenter, show how density waves dissipate
   more quickly in the radiative disk (left plot) than in 
   the isothermal disk (right plot).
   }
   \label{fig15}
   \end{figure}

One limitation of our approach is related to the heavy computational
load needed to complete a simulation. 
Even the 2D simulations performed by \cite{Nelson00} 
are limited to a few binary revolutions. 
The main cause of this is
related to the energy equation solution that
requires a very
short time step owing to the large and short--term variations in
internal energy of the gas at the shock waves and at the location
where material from one disk impacts the other. 
In effect, our models and those presented in \cite{Nelson00} 
are the only ones where each star of the binary system
has its own disk. Simulations of radiative disks with 
grid codes model only the disk around the primary star,
and they do not have to deal with the mass exchange between 
the two disks. 
In SPH simulations it would in theory be possible to speed 
up the computations by implicitly solving the 
energy equation. However, this approach  
fails to 
properly follow the behavior of 
the gas internal energy in the highly perturbed environment
of disks in binaries, as discussed in more detail in the 
next section.
Concerning the long--term evolution of the system we have
studied, 
it is encouraging  that 2D simulations of radiative 
and self--gravitating disks in binaries \citep{muller12,mabash2} 
show stable behavior, that does not evolve significantly
even at later stages. 
This is 
possibly 
related to the fast radiative damping of density waves 
as suggested in \cite{mabash2}. 
The disk 
eccentricity and temperature profiles we obtain in 3D are 
similar to those observed for disks around the primary 
star in the previously mentioned papers, and they are self--similar
after many binary revolutions. 
As a consequence, we may infer that the features 
we find in our simulations, such as the hydraulic jumps, are
preserved during the subsequent evolution of the disks. 
As an additional test, we performed an isothermal simulation 
of the HIDECL model assuming a temperature profile similar
to that of our radiative model at apocenter. After 20 binary revolutions,
the main features due to the binary
perturbations (spiral waves and vertical excursions)
are still  present and have the same morphology at any pericenter passage
during the full length of the
simulation.
However, we have to point out
that there are significant differences between the radiative and 
isothermal runs. As already pointed out by 2D simulations, 
spiral waves are damped more quickly in the radiative model 
as illustrated in the bottom plots of Fig.~\ref{fig15}.
The integrated density distribution of the
primary disks in the radiative (left)  and
isothermal (right) models are compared in the
same orbital configuration after the pericenter
passage. The density waves are significantly less
marked in the radiative case, suggesting that they
are more effectively damped than in the
isothermal one.
The scale height of the disk is higher in the 
radiative case, and the hydraulic jumps have a significantly 
larger jump factor $J_f$ (see Fig.~\ref{fig1}) compared to 
the isothermal 
case illustrated in Fig.~\ref{fig15} top plot. 

From Fig.~\ref{fig15} (top plot) it also appears that the 
density beyond the shock wave is higher in the isothermal 
model. This may be analytically justified when taking into account
that 
in a radiative disk the ratio between the 
pre-- and post--shock densities can be estimated, 
according to \cite{Durisen11}, as 
      \begin{equation}
        \frac{\rho_2}{\rho_1}= \frac{\gamma+1}{\gamma-1}
        \label{rhodu1}
      \end{equation}
for $M >>1$. In an isothermal disk, on the other hand, 
the ratio can be approximated by
      \begin{equation}
        \frac{\rho_2}{\rho_1}= \gamma M^2.
        \label{rhodu2}
      \end{equation}
In this last case, the post--shock density is 
expected to be 
higher than in the radiative case.

\subsection{Future improvements}

\begin{figure}
         \centering
         \includegraphics[width=0.5\textwidth]{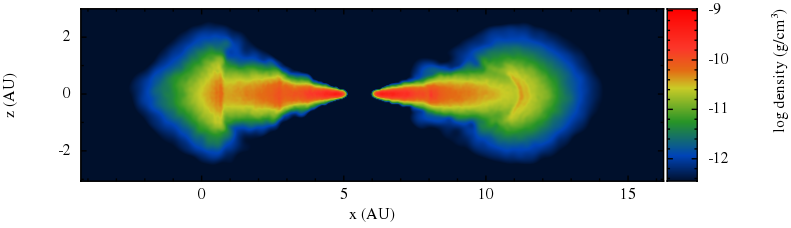}
         \caption{Non--integrated disk density of the 
         primary disk of the HIDECL model
         on the x--z plane. The hybrid method
         proposed by \cite{forgan09} for radiative cooling 
         is used with an explicit time step. 
         }
         \label{fig14}
      \end{figure}

Alternative cooling algorithms can be explored to test their
influence on the physical behavior of the disks. 
We tested the 
cooling algorithm proposed in \cite{stama07} to solve the energy 
equation with an implicit time step with two different 
approaches: using the polytropic cooling term or with the 
hybrid approach proposed by \cite{forgan09} where the cooling 
is coupled to the flux--limited radiation diffusion already 
implemented in our algorithm. In both cases 
we observed the formation of unphysical mass condensations 
in the midplane of the disk after the passage of 
shock waves and significant spikes in the gas temperatures 
all over the disk. This is not due to the cooling and radiative 
transfer algorithms but to the application of the implicit time step. 
Using the same algorithms but with 
an explicit time step for the VINE leapfrog scheme,
both problems disappeared, suggesting that
they were related to the too long time step assumed in the 
implicit method not able to correctly follow the fast 
evolution of the internal energy in our type of system.  
We propagated our standard model HIDECL for a perihelion 
passage with the hybrid method
proposed by \cite{forgan09} without the implicit time step,
and the outcome is very similar to what is obtained with 
the boundary particle algorithm. The only noticeable difference 
is a slightly lower temperature of the gas where  material from 
one disk impacts the other one. In Fig.~\ref{fig14} we show 
a vertical slice of the disk to be compared with that in 
Fig.~\ref{fig1} of the standard model HIDECL with boundary 
particles. Marked hydraulic jumps are visible in both figures, 
and only 
minor differences appear in the vertical 
density distribution.

In our model we neglect stellar irradiation, which might be 
important at least in the outer parts of the disk where the 
viscous heating is less intense. However, it appears a complex
problem to include the stellar flux since 
the hydraulic jumps in the vertical direction
can shield the outer parts of the disk from stellar irradiation. 
Disk self--shadowing
due to spiral waves may strongly reduce the amount of 
stellar irradiation on the disk surfaces. It would be necessary 
to adopt a ray--tracing approach for each particle. This 
appears feasible but might further increase the CPU load. 
In addition, the outer parts of the disks in our ``close'' models
are strongly perturbed by tidal waves and mass exchange whose
effects on the temperature may overcome those due to 
stellar irradiation. 

It would also be interesting to explore the evolution of 
non coplanar disks to test the formation of spiral waves, 
warping, disk precession, and relaxation in this configuration. 
Previous 
SPH models \citep{larw96} were based on the polytropic approximation,
while \cite{franel10} used the isothermal approximation. Even in this 
case, the radiative part of the code might be critical in terms 
of CPU time possibly reducing the model timespan. 

\begin{acknowledgements}
We thank an anonymous referee for useful comments and suggestions. 
Many of our plots were made with the SPLASH software package
\citep{Price07splash}. 
We also thank D. Stamatellos for giving us his tabulated 
pseudo--mean opacities and internal energies. 

\end{acknowledgements}

\bibliography{article}
\bibliographystyle{aa}
\end{document}